\documentclass[12pt]{article}
\usepackage{amsmath,amsfonts}
\pdfoutput=1
\usepackage{hyperref}
\usepackage{graphicx}
\makeatletter\@addtoreset{equation}{section}\makeatother

\DeclareMathOperator{\Tr}{Tr}

\DeclareMathOperator{\arccosh}{arccosh}

\def\bH {\mathbb{H}}
\def\bR {\mathbb{R}}
\def\bZ {\mathbb{Z}}

\newcommand{\be}{\begin{equation}}
\newcommand{\ee}{\end{equation}}
\newcommand{\bea}{\begin{eqnarray}}
\newcommand{\eea}{\end{eqnarray}}

\newcommand{\vev}[1]{{\left< {#1} \right>}}

\newcommand{\eqn}[1]{(\ref{#1})}
\newcommand{\nn}{\nonumber}

\newcommand{\ad}{{\dot a}}
\newcommand{\bd}{{\dot b}}

\newcommand{\cA}{{\mathcal A}}

\newcommand{\cN}{{\mathcal N}}
\newcommand{\cP}{{\mathcal P}}
\newcommand{\cQ}{{\mathcal Q}}

\newcommand{\cS}{{\mathcal S}}

\renewcommand{\title}[1]{\vbox{\center\LARGE
\bf\mathversion{bold}{#1}\mathversion{normal}
}\vspace{5mm}}
\renewcommand{\author}[1]{\vbox{\center#1}\vspace{5mm}}
\newcommand{\address}[1]{\vbox{\center\em#1}}
\newcommand{\email}[1]{\vbox{\center\tt#1}\vspace{5mm}}

\begin{document}
\begin{titlepage}
\begin{center}
\vspace{5mm}
\hfill {\tt HU-EP-09/05}\\
\vspace{20mm}
\title{BPS Wilson loops in $\cN=4$ SYM: Examples on hyperbolic submanifolds of 
space-time}

\author{\large Volker Branding$^{1,2}$ and Nadav Drukker$^2$}

\address{
$^1$Institut f\"ur Mathematik, Universit\"at Potsdam\\
Am Neuen Palais 10, 14469 Potsdam\\
\smallskip
$^2$Institut f\"ur Physik, Humboldt-Universit\"at zu Berlin,\\
Newtonstra{\ss}e 15, D-12489 Berlin, Germany}

\email{volker.branding@gmail.com,\\
drukker@physik.hu-berlin.de}

\end{center}

\abstract{
\noindent
In this paper we present a family of supersymmetric Wilson loops of 
$\cN=4$ supersymmetric Yang-Mills theory in Minkowski space. Our examples 
focus on curves restricted to hyperbolic submanifolds, $\bH_3$ and 
$\bH_2$, of space-time. Generically they preserve two supercharges, but 
in special cases more, including a case which has not been discussed 
before, of the hyperbolic line, conformal to the straight line and circle, which 
is half-BPS. 
We discuss some general properties of these Wilson loops and 
their string duals and study special examples in more detail. Generically 
the string duals propagate on a complexification of $AdS_5\times S^5$ 
and in some specific examples the compact sphere is effectively 
replaced by a de-Sitter space.}

\vfill

\end{titlepage}

{\addtolength{\parskip}{-1ex}\tableofcontents}

\section{Introduction and summary}

The spectrum of operators in $\cN=4$ supersymmetric Yang-Mills (SYM) 
which preserve some supersymmetry 
is very rich. The list of operators which are half-BPS is still not 
complete,\footnote{For recent progress see \cite{D'Hoker:2008ix}.}
and of course much less is known about operators preserving 
fewer supercharges. The dynamics of a supersymmetric operator is much more 
constrained than that of a generic object and the more supercharges it preserves, 
the more constrained it is. Therefore an interesting endeavor is to 
find operators preserving relatively few supercharges, whose dynamics are 
rather rich, but may still be under control.

An example of such operators are the BPS Wilson loops constructed 
in \cite{DGRT-more,DGRT-YM2,DGRT-big}. For an arbitrary 
curve on an $S^3$ submanifold of $\bR^4$ a prescription is given for 
choosing extra scalar coupling such that the resulting Wilson loop preserves 
at least two supercharges. When the loop is on $S^2$ it preserves four and 
in some special examples, eight, while for the circle it preserves sixteen.

An earlier construction of Zarembo 
\cite{Zarembo:2002an} does much the same with arbitrary curves on 
linear subspaces of $\bR^4$. In that case the expectation value of all the 
loops is unity, even if they preserve only a single supercharge 
\cite{Guralnik:2003di,Guralnik:2004yc,Dymarsky:2006ve}. 
The interesting thing about the Wilson loops on $S^3$ is that their 
expectation value is not so simple. In the case of the circle it 
is given by a zero dimensional matrix model 
\cite{Erickson,Dru-Gross}
as was recently proven by Pestun \cite{Pestun:2007rz}.

More remarkably, in \cite{DGRT-YM2,DGRT-big} some evidence was 
presented that this result extends to arbitrary loops on $S^2$, preserving 
four supercharges. In that case the propagators are not constant in the 
Feynman gauge, but the final result seems to be the same matrix model, 
with a modified coupling. This result can be motivated, as an intermediate 
step, by a perturbative calculation in 2-dimensional bosonic Yang-Mills (YM) 
theory, an observation which was confirmed at the two-loop order 
(for single loops) in \cite{Dru-1/4,Bassetto:2008yf,Young:2008ed}.

In this paper we study another family of supersymmetric Wilson loops, this 
time in Minkowski space $\bR^{3,1}$. Working with indefinite metric 
expands the range of possibilities. For one, there are Euclidean 
two-spheres in Minkowski space, so the previously mentioned results would apply 
there too, but there are no three-spheres. Instead the maximally symmetric 
hypersurfaces are the light-cone, the Lorentzian hyperboloid (de-Sitter space) 
and the Euclidean hyperboloid $\bH_3$ (Euclidean $AdS_3$).

Light-like Wilson loops, even 
without coupling to the scalar fields, are supersymmetric. 
More generally, we expect constructions like that of Zarembo \cite{Zarembo:2002an} 
would work also on Minkowski space, as was recently utilized in 
\cite{Ishizeki:2008dw}.

In this paper we study the case of $\bH_3$. For a general curve constrained 
to this submanifold we find a prescription to choose the scalar couplings 
such that the resulting loop is BPS. We then focus on specific subclasses 
of examples preserving more supersymmetry, in particular $\bH_2$ and 
special loops along curves of constant curvature. We find results mostly in 
parallel with those on $S^3$ and its submanifolds. Yet there are some 
differences that are not so trivial, specifically in understanding the 
string-theory dual --- where an analytic continuation has to be performed 
on an $S^2$ subspace of $S^5$ which yields a two-dimensional de-Sitter 
space $dS_2$.

One particular reason to study Wilson loops in Minkowski space is 
the connection between Wilson loops and scattering 
amplitudes. A great deal of evidence shows that Wilson 
loops made out of light-like segments 
calculate maximum helicity violating gluon scattering 
amplitudes \cite{Alday:2007hr, Drummond:2007aua, 
Brandhuber:2007yx, Bern:2008ap, Drummond:2008aq} 
(for a review, see \cite{Alday:2008yw}). 
It is therefore also natural to try to find supersymmetric Wilson loops 
in Minkowski space.

We present the construction of the Wilson loops on $\bH_3$ in 
Section~\ref{sec-general} and prove that they are supersymmetric. Then 
we discuss the restriction to $\bH_2$, show how the connection to 
2-dimensional Yang-Mills shows up again at the leading order in perturbation 
theory. This connection suggests once more that the expectation value of 
these Wilson loops is given by a Gaussian matrix model, at least 
for compact curves.%
\footnote{Unlike the sphere the Hyperboloid is not compact, 
so one can construct Wilson loop operators with an infinite extent, some 
examples of which are presented in Section~\ref{sec-examples}.}
We then discuss some general properties of the string duals of these 
Wilson loops.

The construction of BPS Wilson loops in $\cN=4$ SYM always involves 
extra couplings to the scalar fields. For curves on $S^3$ these couplings 
can be all real, but for our construction here they will generically be 
complex. As the scalar couplings translate in the string dual to positions 
on $S^5$, we are forced to consider the embedding of the string into a 
complexification of $AdS_5\times S^5$. This is discussed in 
Section~\ref{subsec-dS2}.

In Section~\ref{sec-examples} we present some special examples. The first 
is the hyperbolic line, which is $1/2$-BPS, like the line and circle. 
Indeed it is related to them by a conformal transformation. After 
that, we present three other examples of circles, other hyperbolic lines 
and cusps which are all $1/4$ BPS and in all cases we can find the 
corresponding string solutions. Two of the examples 
require a complexification of the target space, so 
the string can be viewed as propagating in a de-Sitter space rather than 
on an $S^2$. In the one case where the loop is compact, the $1/4$ BPS 
circle, we can interpolate the gauge theory result to strong coupling by 
summing over ladder diagrams and find complete agreement.

Some of the results contained in the paper are a summary of the diploma 
thesis \cite{Volker-thesis}.

\section{Generalities}
\label{sec-general}

\subsection{Construction}
\label{subsec-construction}

We consider Wilson loops confined to an $\bH_3$ subspace of
Minkowski space $\bR^{3,1}$ given by
\begin{equation}
-x_0^2+x_1^2+x_2^2+x_3^2=-1\,,\qquad
x_0>0\,.
\end{equation}
This is a Euclidean manifold contained within the future light-cone.

Our construction is based on a Wick-rotation of the Wilson loops on
$S^3$ \cite{DGRT-more,DGRT-big} where an important
ingredient were the invariant one-forms. They are Wick-rotated to
\begin{equation}
\begin{aligned}
\omega_1&=x^0dx^1-x^1dx^0+i(x^2dx^3-x^3dx^2)\,,\\
\omega_2&=x^0dx^2-x^2dx^0+i(x^3dx^1-x^1dx^3)\,,\\
\omega_3&=x^0dx^3-x^3dx^0+i(x^1dx^2-x^2dx^1)\,.
\end{aligned}
\label{one-forms}
\end{equation}
We allow the Wilson loops to follow an arbitrary path on this manifold
and in order to preserve supersymmetry will couple them to three
of the real scalar fields, which we take to be $\Phi^1$, $\Phi^2$
and $\Phi^3$. The coupling can be expressed with the use of the
one-forms (\ref{one-forms}) in terms of the modified connection
($A=A_\mu dx^\mu$ is the gauge connection)
\begin{equation}
\tilde{A}=A-i\,\omega_i\Phi^i\,,
\end{equation}
as
\begin{equation}
W=\frac{1}{N}\Tr\cP\exp\oint i\tilde{A}\,.
\label{WL}
\end{equation}

\subsection{Supersymmetry}
\label{subsec-susy}

To check that these Wilson loops are indeed BPS objects, consider
the supersymmetry variation which is proportional to
\begin{equation}
\delta_\epsilon W\propto
(\gamma_\mu dx^\mu-i\rho^i\gamma^5\omega_i)\epsilon(x)\,.
\label{susy}
\end{equation}
Here $\gamma_\mu$ are the usual Dirac matrices, $\rho^i$ are
three of the gamma matrices of $SO(6)$ and we take them to
commute with $\gamma_\mu$. $\epsilon(x)$ is the conformal
Killing spinor in flat space given by two constant spinors as
\begin{equation}
\epsilon(x)=\epsilon_0+x^\mu\gamma_\mu\epsilon_1\,.
\end{equation}
$\epsilon_0$ is the parameter for the action of the Poincar\'e
supercarges ($Q$'s) and $\epsilon_1$ that for the action of the
superconformal charges ($S$'s).

From the definition of the one-forms (\ref{one-forms}) one can see that 
they are related to the generators of
the Lorentz group in the chiral spinor representation ($\tau^i_L$
are the Pauli matrices acting on the chiral spinors)
\begin{equation}
\omega_i\tau^i_L=x^\mu dx^\nu\gamma^+_{\mu\nu}\,.
\end{equation}
The supersymmetry variation (\ref{susy}) should vanish at all
points on $\bH_3$ and for arbitrary tangent vectors. We note
that in a chiral basis the Dirac matrices are represented by Pauli matrices
\begin{equation}
\gamma^i\epsilon^\pm=\pm i \tau^i\epsilon^\pm\,,\qquad
\gamma^0\epsilon^\pm=i\epsilon^\pm\,.
\end{equation}
If we restrict to chiral spinors, the BPS condition reduces to
\begin{equation}
\omega_i\left[-i(\rho^i\epsilon_0^+-i\tau^i_L\epsilon_1^+)
-x^\rho\gamma_\rho(\tau^i_L\epsilon_0^+-i\rho^i\epsilon_1^+)\right]
=0\,,
\end{equation}
so we get the three independent conditions
\begin{equation}
\rho^i\epsilon_0^+=i\tau^i_L\epsilon_1^+\,,\qquad
i=1,2,3\,.
\label{ep0-ep1}
\end{equation}
To solve these equations one can eliminate $\epsilon_1^+$ and get
the equations for $\epsilon_0^+$
\begin{equation}
\rho^{ij}\epsilon_0^+=-\tau^i_L\tau^j_L\epsilon_0^+
=-i\varepsilon^{ijk}\tau^k_L\epsilon_0^+\,.
\end{equation}
This is a set of three equations, but only two are independent.

Now we notice that $\rho^{ij}$ are the generators of
a subgroup of the $SU(4)$ $R$-symmetry group, which we label
$SU(2)_A$. The indices for the $\bf4$ of $SU(4)$ are split into a pair 
$\ad a$ for $SU(2)_A\times SU(2)_B$ respectively, where the second 
group acts on the three remaining scalar fields. $\rho^{ij}$ can then 
be represented by pauli matrices
$\rho^{ij}=i\varepsilon^{ijk}\tau^k_A$. This leads to the three
equations
\begin{equation}
(\tau^i_A+\tau^i_L)\epsilon_{0\ad a}^{+\,\alpha}=0\,,\qquad
i=1,2,3\,.
\end{equation}
This means that $\epsilon_0^+$ should be a singlet of the diagonal sum of 
the $SU(2)_A$ and $SU(2)_L$ groups. We can express it in terms 
of an arbitrary 2-component spinor $\epsilon_a$ as
\begin{equation}
\epsilon_{0\ad a}^{+\,\alpha}=
(\delta^\alpha_1\delta_{\ad}^2-\delta^\alpha_2\delta_{\ad}^1)\epsilon_a=
i(\tau^2)^\alpha{}_{\ad}\epsilon_a\,.
\end{equation}

Now using (\ref{ep0-ep1}) we derive%
\footnote{Writing this requires lowering the indices of $\epsilon_1^+$, which 
corresponds to a specific representation of the $\rho^i$ matrices.}
\begin{equation}
\epsilon_{1\alpha}^{+\ad a}
=-i\varepsilon_{\alpha\beta}\varepsilon^{\ad\bd}\varepsilon^{ab}
(\tau^1_A)_{\dot b}{}^{\dot c}(\tau^1_L)^\beta{}_\gamma\,\epsilon_{0\dot c b}^{+\,\gamma}
=(\tau^2)^{\ad}{}_\alpha\varepsilon^{ab}\epsilon_b\,,
\end{equation}
From this we finally see that the general Wilson loop on $\bH_3$ of the type 
\eqn{WL} will be invariant under the two supercharges
\be
\cQ^a=
i(\tau^2)^\alpha{}_{\ad}\,Q_\alpha^{\ \ad a}
+(\tau^2)^{\ad}{}_\alpha\,\varepsilon^{ab}\,S^\alpha_{\ \ad b}\,.
\ee

\subsection{Restriction to $\bH_2$}
\label{subsec-H2}
A simple restriction on the possible curves is given by setting
$x_3=0$, resulting in an $\bH_2$ subspace. In this case the scalar couplings
are given by
\begin{equation}
\omega_i=(x^0dx^1-x^1dx^0,\,x^0dx^2-x^2dx^0,\,
i(x^1dx^2-x^2dx^1))\,.
\label{H2-one-forms}
\end{equation}
This is clearly the analog of the $S^2$ loops discussed in 
\cite{DGRT-more,DGRT-YM2,DGRT-big}

A simple calculation shows that for a general path on this
space these scalar couplings will guarantee that the loop preserves
four supercharges. In addition to the chiral supercharges, these loops 
also preserve two anti-chiral ones. 
In the case of $S^2$ it was shown \cite{DGRT-YM2,DGRT-big} 
that at leading order in perturbation theory the calculation of the 
supersymmetric Wilson loops is related to that of loops 
in 2-dimensional Yang-Mills. In special cases this was checked also to 
2-loop order \cite{Bassetto:2008yf,Young:2008ed}, and in more restricted 
cases there are tests that go even beyond that 
\cite{Dru-1/4, DGRT-1/4-giant}. We show in the next subsection that this 
result extends, at least at leading order, to the loops on $\bH_2$.

Note that by restricting $x_0$ to be a constant one would end
with an $S^2\subset\bH_3$. Unlike the case of $\bH_2$, a general
curve on this subspace does not preserve extra supersymmety, so
we will not consider it in detail.

In Section~\ref{sec-examples} we present special examples of
Wilson loops that preserve more than these 4 supercharges. They
will all be constrained within this subspace.

\subsection{Perturbative calculation: 2-dimensional Yang-Mills
on $\bH_2$ and the matrix model}
\label{subsec-2dYM}

We can write $\bH_2$ embedded in $\bR^{2,1}$ in terms of the complex
coordinates $\zeta$ and $\bar\zeta$ in the unit disc as
\begin{equation}
x_0=\frac{1+\zeta\bar\zeta}{1-\zeta\bar\zeta}\,,\qquad
x_1=\frac{\zeta+\bar\zeta}{1-\zeta\bar\zeta}\,,\qquad
x_2=-i\,\frac{\zeta-\bar\zeta}{1-\zeta\bar\zeta}\,.
\label{stereo}
\end{equation}
This gives the metric on the Poincar\'e disc
\begin{equation}
ds^2=\frac{4\,d\zeta\,d\bar{\zeta}}{(1-\zeta\bar{\zeta})^2}\,.
\label{metric-poincare}
\end{equation}
Now we consider the perturbative expansion of the Wilson loop and write down 
the propagator between two points along the curve,
one at $x$ represented by $(\zeta,\bar\zeta)$ and one at
$x'$ represented by $(\eta,\bar{\eta})$. Working in the Feynman
gauge the effective propagator combining the gauge fields
and scalars is
\begin{equation}
\begin{aligned}
\vev{(iA_\mu^a dx^\mu+\Phi^{i\,a}\omega_i)
(iA_\mu^b dx'^\mu+\Phi^{i\,b}\omega'_i)}
&=
-\frac{g_{4d}^2\,\delta^{ab}}{4\pi^2}\,
\frac{dx\cdot dx'-\omega_i\omega_i'}
{(x-x')^2}
\\&\hskip-1.5in=
-\frac{g_{4d}^2\,\delta^{ab}}{4\pi^2(1-\zeta\bar\zeta)(1-\eta\bar{\eta})}
\left[\frac{\bar\zeta-\bar{\eta}}{\zeta-\eta}\,d\zeta\,d\eta
+\frac{\zeta-\eta}{\bar\zeta-\bar{\eta}}\,d\bar{\zeta}\,d\bar{\eta}\right].
\end{aligned}
\label{one-loop}
\end{equation}

We note, that as in the case of the loops on $S^2$ 
\cite{DGRT-YM2,DGRT-big}, this is
a propagator for a gauge field in two dimensions. Consider YM$_2$
on $\bH_2$ in the generalized Feynman gauge
\begin{equation}
L=\frac{1}{g_{2d}^2}\left[\frac{1}{4}(F_{\alpha\beta}^a)^2
+\frac{1}{2\xi}(\nabla^\alpha A_{a\,\alpha})^2
+\partial_\alpha b^a(D^\alpha c)^a\right]\,,
\end{equation}
where
\begin{equation}
F_{\alpha\beta}^a=\partial_\alpha A^a_\beta
-\partial_\beta A^a_\alpha+f^{abc}A_\alpha^bA_\beta^c\,,
\qquad
(D_\alpha c)^a=\partial_\alpha c^a+f^{abc}A_\alpha^bc^c\,,
\end{equation}
and $\nabla^\alpha$ is the covariant derivative with respect to the
metric (\ref{metric-poincare}). Setting $\xi=-1$
the gauge term becomes
\begin{equation}
\sqrt{g}\,L=
-\frac{(1-\zeta\bar{\zeta})^2}{2g_{2d}^2}
\left[(\partial_\zeta A_{\bar\zeta})^2+
(\partial_{\bar\zeta}A_\zeta)^2\right].
\end{equation}
It is easy to check that the propagators
\begin{equation}
\begin{aligned}
\Delta_{\zeta\zeta}^{ab}(\zeta,\bar\zeta;\eta,\bar{\eta})
&=\frac{g_{2d}^2\,\delta^{ab}}{\pi(1-\zeta\bar\zeta)(1-\eta\bar{\eta})}
\frac{\bar\zeta-\bar{\eta}}{\zeta-\eta}\,,\\
\Delta_{\bar\zeta\bar\zeta}^{ab}(\zeta,\bar\zeta;\eta,\bar{\eta})
&=\frac{g_{2d}^2\,\delta^{ab}}{\pi(1-\zeta\bar\zeta)(1-\eta\bar{\eta})}
\frac{\zeta-\eta}{\bar\zeta-\bar{\eta}}\,,\\
\end{aligned}
\label{2d-propagators}
\end{equation}
satisfy
\begin{equation}
\frac{1}{2g_{2d}^2}\,\partial_{\bar\zeta}\left[(1-\zeta\bar\zeta)^2
\partial_{\bar\zeta}\Delta_{\zeta\zeta}^{ab}(\zeta,\bar\zeta;\eta,\bar{\eta})\right]
=\delta^{ab}\delta^2(\zeta-\eta)\,,
\end{equation}
and likewise for $\Delta_{\bar\zeta\bar\zeta}^{ab}$.

As was discussed in the $S^2$ case in \cite{DGRT-big},
these propagators use the Wu-Mandelstam-Leibbrandt prescription
for dealing with singularities in the Euclidean light cone
propagator \cite{Wu:1977hi,Mandelstam:1982cb,Leibbrandt:1983pj}.
This prescription leads to different results than
one would get by a Wick-rotation from Lorentzian signature.
This difference was stressed in
\cite{Staudacher:1997kn} and resolved in
\cite{Bassetto:1998sr}, where the authors realized that this
prescription gives the same result as one gets by doing the
instanton expansion of \cite{Witten:1992xu} and focusing on the
zero-instanton sector only.

Now consider an arbitrary Wilson loop of 2-dimensional YM on
$\bH_2$
\begin{equation}
W_{2d}=\frac{1}{N}\,\Tr\cP\exp\oint i(A_\zeta\,d\zeta+A_{\bar\zeta}\,d\bar\zeta)\,.
\end{equation}
At leading order in perturbation theory, using the propagators
(\ref{2d-propagators}) this loop will agree precisely with the
expression for the Wilson loop in four dimensions (\ref{one-loop})
under the identification of the couplings
\begin{equation}
g_{2d}^2=\frac{g_{4d}^2}{4\pi}\,.
\end{equation}

We conclude that as in the case of the BPS loops on $S^2$,
the Wilson loops on $\bH_2$ seem to be given by this perturbative
calculation in two dimensions, at least at the one-loop level.
In the case of $S^2$ the relation between the couplings was
such that $g_{2d}^2$ was negative, but here it is positive.

If the loop is compact and encloses a region of area $\cA$, it is easy to 
evaluate the result of the YM$_2$ calculation. Following 
\cite{Staudacher:1997kn} it is possible to use the invariance under 
area-preserving diffeomorphisms and do the explicit calculation in the 
case of the circle. At leading order we get
\be
\vev{W}_{2d}=1-g_{2d}^2N\,\frac{\cA(\cA+4\pi)}{8\pi}\,.
\ee
The full perturbative series can be expressed in terms of Laguerre 
polynomials \cite{Staudacher:1997kn,Dru-Gross}
\be
\vev{W}_{2d}
=\frac{1}{N}\,L_{N-1}^1\left(g_{2d}^2\,\frac{\cA(\cA+4\pi)}{4\pi}\right)
\exp\left(-g_{2d}^2\,\frac{\cA(\cA+4\pi)}{8\pi}\right).
\ee
Assuming this result is exact in the four dimensional theory too and 
concentrating on the large $N$ limit, the result reduces to a Bessel function
\be
\vev{W}_\text{planar}=
\frac{4\pi}{\sqrt{g_{4d}^2N\cA(\cA+4\pi)}}\,
J_1\left(\frac{\sqrt{g_{4d}^2N\cA(\cA+4\pi)}}{2\pi}\right)\,.
\label{bessel}
\ee
In the case of the loops on $S^2$ there were some very interesting 2-loop 
calculations by Bassetto et al. and Young
\cite{Bassetto:2008yf,Young:2008ed}. It would be interesting to 
repeat those checks for this case.

This result is identical to that of the $1/2$ BPS circle 
\cite{Erickson,Dru-Gross,Pestun:2007rz} under the replacement 
$g_{4d}^2N\to-g_{4d}^2N\cA(\cA+4\pi)/4\pi^2$. This expression 
is also equal to that of the Wilson loop in the Gaussian matrix model. 
One difference with respect to the circle is the change in sign on 
the coupling, which has the effect of replacing the modified Bessel 
function $I_1$ with the regular one $J_1$. This function has 
oscillatory behavior at large $\lambda$, where the Wilson loop 
is described by a string, suggesting that the string dual should be 
Lorentzian.

Another difference from the circle, or more generally the loops on $S^2$, 
is that on $\bH_2$ there are non compact loops that go off to infinity. Such 
loops are interesting since they asymptote to the light-cone and are 
therefore similar to light-like cusped Wilson loops that have been used 
to calculate scattering amplitudes. In these cases it is unclear how 
the reduction to YM$_2$ would work. It is still true that the effective 
propagator is the same as in the lower dimensional theory, but 
a non-compact curve is not completely gauge invariant. Large gauge 
transformations will change its expectation value and therefore one cannot 
rely on the gauge invariance of the two-dimensional theory and perform 
the calculation in the light-cone gauge. It is therefore left as a question 
whether the non compact loops are also captured beyond the leading 
level by two-dimensional YM in some gauge.

\subsection{String theory description: Complexification and de-Sitter space}
\label{subsec-dS2}

We now discuss some of the basic properties of the string
theory duals \cite{Rey-Yee,Maldacena-wl} of the BPS Wilson loops
on $\bH_3$. The issue we would like to address is that for these loops
the scalar couplings are in general imaginary, hence the strings
describing the loops will live on a complexified $S^5$. The most 
general Wilson loop we constructed has complex couplings to 
three scalars, so only three complex directions inside the complexified 
$S^5$ are turned on.

A general requirement for a Wilson loop to be BPS is that the
strength of coupling to the scalars is the same as that to the
gauge fields. Viewing the gauge theory as dimensionally reduced
from $\cN=1$ in ten dimensions, the loop in ten dimensions has
to be light-like.

Therefore one usually considers a real coupling for a space-like
Wilson loop, imaginary coupling for a time-like one and no scalar
couplings at all for a light-like loop in 4-dimensions. These loops
then have a natural description in the $AdS_5\times S^5$ dual of the
gauge theory.

But in general one may consider complex scalar couplings, where now
the BPS-condition is that they are light-like in a complexification
of (six of the coordinates of) the ten-dimensional space. Indeed in
our construction (\ref{one-forms}) the scalar couplings are complex.

In such situations the dual $AdS$ interpretation is rather subtle,
and requires also complexification. A related example is in the
study of charged local operators in the Euclidean gauge theory.
The charged local operators (like the BMN ground state $\Tr Z^J$ 
\cite{BMN})
can be described semiclassically in the Lorentzian theory by
particle trajectories or giant gravitons. In the Euclidean
theory, there is no real time and therefore one cannot describe a
real process of propagating from the boundary of $AdS_5$ into the
bulk. This can be resolved by considering a tunneling picture,
or a complexification of the space, which essentially Wick-rotates
one of the directions on $S^5$ (see the discussion in 
\cite{Yoneya:2003mu}).

In the case of the Wilson loops on $S^3$ coupled to three scalars, 
the dual string is on an $\bH_4\times S^2$ subspace of 
$AdS_5\times S^5$. This subspace can be represented 
in terms of the coordinates
\be
ds^2=\frac{1}{z^2}\left(dz^2+dx_1^2+dx_2^2+dx_3^2+dx_4^2\right)
+dy_1^2+dy_2^2+dy_3^2\,,
\label{S2-metric}
\ee
subject to the constraints
\be
z^2+x_1^2+x_2^2+x_3^2+x_4^2=1\,,\qquad
y_1^2+y_2^2+y_3^2=1\,.
\ee
The first condition is the extension into the bulk of the condition of being on 
$S^3$ on the boundary, while the second is that of $S^2\subset\bR^3$.

We want to modify this setup for the case at hand, of the Wilson loops on 
$\bH_3$. Now some of the scalar couplings may be complex, and hence the 
$y_i$ coordinates too. We consider in detail only loops on 
$\bH_2$, for which only one of the three scalar couplings becomes 
imaginary while the others remain real. Wick-rotating $y_3=i y_0$ and 
$x_4=ix_0$ and setting $x_3=0$ we get
\be
\begin{gathered}
ds^2=\frac{1}{z^2}\left(dz^2-dx_0^2+dx_1^2+dx_2^2\right)
-dy_0^2+dy_1^2+dy_2^2\,,\\
x_0^2-z^2-x_1^2-x_2^2=1\,,\qquad\qquad
-y_0^2+y_1^2+y_2^2=1\,.
\end{gathered}
\ee
Now the $y_i$ coordinates parameterize de-Sitter space $dS_2$ with 
Lorentzian signature. Any non-trivial string embedding into this space 
will be of a string with Lorentzian world-sheet and hence the string 
should propagate also on a Lorentzian submanifold of $AdS_4$. Alas, 
the extra condition on the $x_i$ and $z$ coordinates 
imply that the string moves in an $\bH_3$ subspace of $AdS_4$, which 
is a Euclidean submanifold.

To resolve this issue we propose to Wick-rotate instead of $x_0$, both 
$x_1$ and $x_2$ (which is very similar to rotating $x_0$ and $z$). 
This gives
\be
\begin{gathered}
ds^2=\frac{1}{z^2}\left(dz^2+dx_0^2-dx_1^2-dx_2^2\right)
-dy_0^2+dy_1^2+dy_2^2\,,\\
x_0^2+z^2-x_1^2-x_2^2=1\,,\qquad\qquad
-y_0^2+y_1^2+y_2^2=1\,.
\end{gathered}
\label{dS-metric}
\ee
Now $x_i$ and $z$ parametrize the three-dimensional de-Sitter space $dS_3$. 
To see that define the embedding coordinates
\be
X_0=\frac{x_0}{z}\,,\qquad
X_1=\frac{x_1}{z}\,,\qquad
X_2=\frac{x_2}{z}\,,\qquad
X_3=\frac{1}{z}\,,
\ee
The metric \eqn{dS-metric} reduces to the flat metric for these coordinates with 
signature $(+,-,-,-)$ and the constraint
\be
X_0^2-X_1^2-X_2^2-X_3^2=-1\,.
\ee
In this space we can embed strings with Lorentzian world-sheets.

In attempts to quantize gravity on de-Sitter space its similarity to $AdS$ space is 
often employed 
(see e.g. \cite{Strominger:2001pn, Balasubramanian:2001nb,Maldacena:2002vr}). 
This similarity is realized here by the fact that the target space of 
our $\sigma$-model seems to be automatically continued from $AdS_3\times S^2$ 
to $dS_3\times dS_2$. We will refrain, however, from trying to interpret 
our results as a holographic realization of de-Sitter space and view it rather 
as an analytical continuation.

Yet, for practical purposes we are dealing with de-Sitter space which raises 
some of the usual issues associated to that space, like the choice of global 
structure. Global de-Sitter space has two conformal boundaries, 
one at asymptotic past and one at asymptotic future. An alternative 
to that exists, where antipodal points are identified, giving $dS/\bZ_2$ 
\cite{Schroedinger,Parikh:2002py}. Taking this 
choice in our case has the effect of limiting the range of $y_0$ to $\bR_+$ and 
likewise for $X_0$. Since we got these spaces by a Wick-rotation, it is not 
entirely clear what the range of the coordinates is. While we do not have strong 
arguments against global de-Sitter space, it seems in the examples we study 
below that the interpretation in terms of the orbifold $dS/\bZ_2$ is more natural.

\section{Examples}
\label{sec-examples}

\subsection{$1/2$ BPS hyperbolic line}
\label{subsec-line}

\begin{figure}
\begin{center}
\includegraphics[width=70mm]{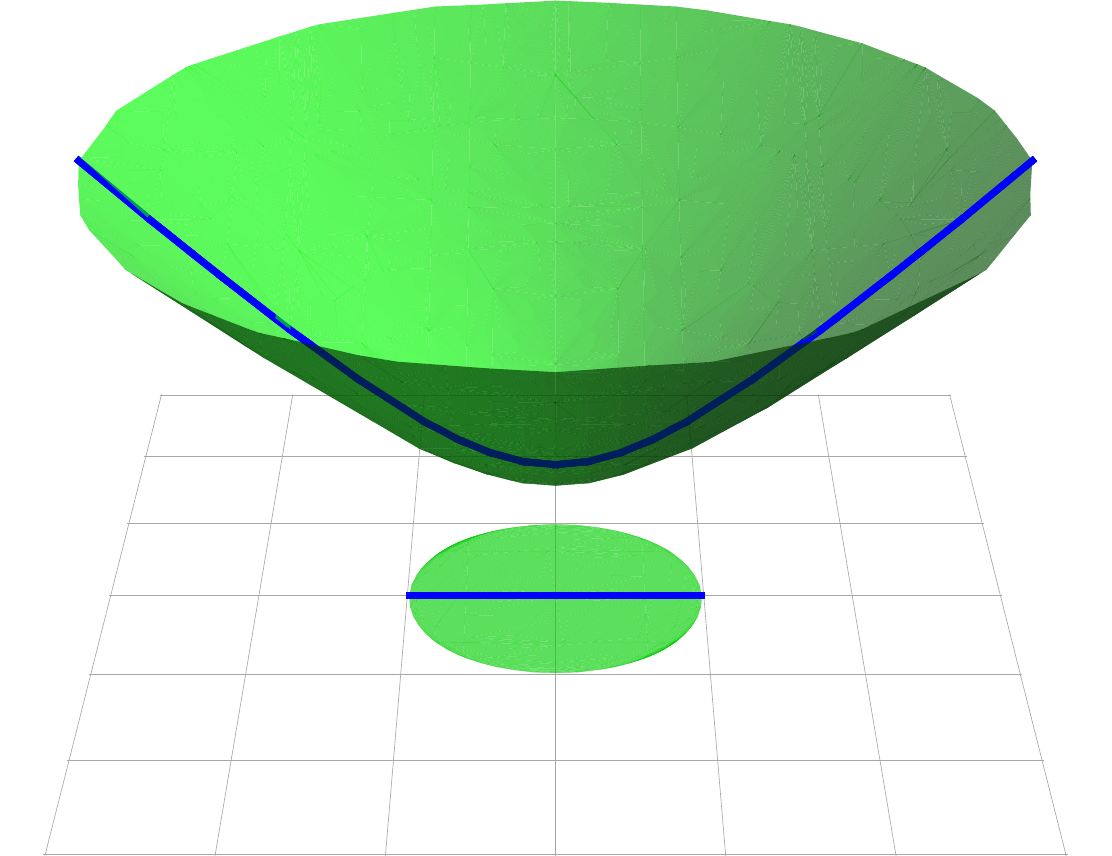}
\parbox{13cm}{
\caption{The hyperbolic line $x_2=0$ represented on the hyperboloid 
$x_0^2-x_1^2-x_2^2=1$ and its stereographic projection onto 
the Poincar\'e disc.
\label{line-fig}}}
\end{center}
\end{figure}

Consider the hyperbolic line
\begin{equation}
x^\mu=(\cosh t,\,\sinh t,\,0,\,0)\,,\qquad
-\infty<t<\infty\,.
\label{hyper-line}
\end{equation}
We can immediately read from (\ref{one-forms}) that the three
scalar couplings will be given by
\begin{equation}
\omega_i=(1,\,0,\,0)\,dt\,,
\end{equation}
so it has constant coupling to a single scalar $\Phi^1$. Checking
the supersymmetry variation of this loop one finds that it is
annihilated by half the supercharges, just like the straight line
or circle.

One may be quite surprised by this, it is generally assumed that
the only half-BPS Wilson loops are the straight line and the circle,
which are related to each other by a conformal transformation. That
is indeed true in Euclidean $\bR^4$, but on Minkowski space there
are some more possibilities: The space-like hyperbolic line like
(\ref{hyper-line}) is clearly a boost of the line in the $x^1$
direction and likewise there is the time-like hyperbolic line
\begin{equation}
x^\mu=(\sinh t,\,\cosh t,\,0,\,0)\,.
\end{equation}
This curve is conformal to a time-like straight line and with
a constant (imaginary) scalar coupling will also be $1/2$ BPS.

In fact this second line belongs to another family of BPS loops
that may be constructed on a 3-dimensional de-Sitter subspace
of Minkowski space. We will not study them in this paper.

Using the representation \eqn{stereo}, the Wilson loop is given by 
$\zeta=\tanh\frac{t}{2}$. Then expanding the loop to second order and 
using \eqn{one-loop} we find
\be
\vev{W}=1-
\frac{g_{4d}^2N}{2}\frac{1}{2}\frac{1}{4\pi^2}
\int dt_1\,dt_2
\frac{2\dot\zeta(t_1)\dot\zeta(t_2)}{(1-\zeta(t_1)^2)(1-\zeta(t_2)^2)}+\cdots
\ee
As for the case of the circle \cite{Erickson}, the integrand, which 
is the combined propagator, is a constant ($1/2$) thus
\be
\vev{W}=1-
\frac{\lambda}{32\pi^2}\int dt_1\,dt_2+\cdots
=1-\frac{\lambda}{32\pi^2}(2T)^2+\cdots
\label{1-loop-1}
\ee
where $T$ is a cutoff on the parameter $t$, corresponding to taking the 
two endpoints
\be
(x_0,x_1)=(\cosh T,-\sinh T)\,,
\qquad
(x'_0,x'_1)=(\cosh T,\sinh T)\,.
\ee
Unlike the circle, this Wilson loop suffers from infra-red divergences, which 
occur also in some of the other examples of Wilson loops listed below. 
These divergences have to do with the fact that the curves 
are not compact and their physical meaning will be studied 
elsewhere.

\begin{figure}
\begin{center}
\includegraphics[width=70mm]{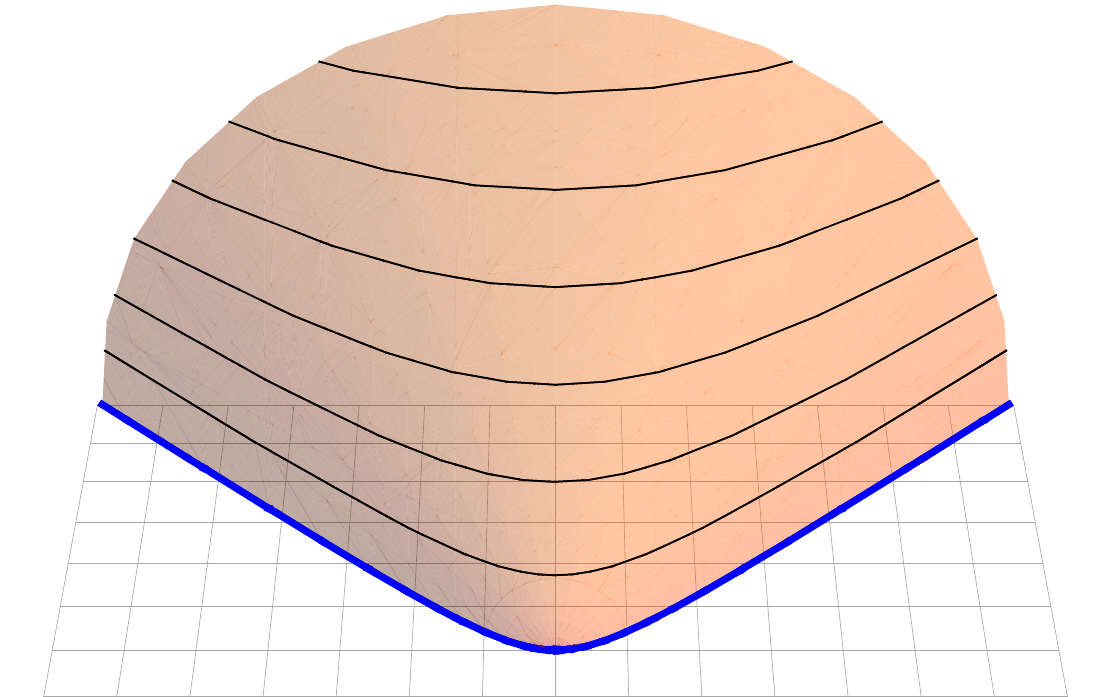}
\parbox{13cm}{
\caption{
A representation of the minimal surface for the hyperbolic line. 
The vertical direction represents the ``$AdS$ direction'' 
$z=\sqrt{x_0^2-x_1^2-1}$. The blue line is at the boundary and 
the thin lines at constant $z$ are a natural choice for a cutoff on 
the world-sheet.
\label{linesurf-fig}}}
\end{center}
\end{figure}

We turn now to finding the string dual to this Wilson loop.
The loop only couples to a single scalar $\Phi^1$ and has a real 
coupling. Therefore we do not need to consider the embedding of the 
string into a complexified $AdS_5\times S^5$. Rather the string is 
localized at a constant point
$y_2=y_3=0$ on $S^5$. For the $AdS_5$ solution we consider an
$AdS_3$ subspace with metric
\begin{equation}
ds^2=\frac{1}{z^2}(dz^2-dx_0^2+dx_1^2)\,.
\label{AdS-metric}
\end{equation}
The string surface should end along the curve $x_0^2-x_1^2=1$ on
the boundary at $z=0$. From symmetry considerations (and since
it is conformal to the straight line and to the circle) we know that
the string will span an $\bH_2$ subspace of target space which is
given by the equation
\begin{equation}
z^2=x_0^2-x_1^2-1\,.
\label{line-constraint}
\end{equation}

This can be verified, of course, by checking the equations of motion.
The simplest way to write them is to make a change of coordinates
\begin{equation}
x_0=e^w\cosh\alpha\cosh t\,,\qquad
x_1=e^w\cosh\alpha\sinh t\,,\qquad
z=e^w\sinh\alpha\,,
\label{line-coords}
\end{equation}
which gives the metric
\begin{equation}
ds^2=\frac{1}{\sinh^2\alpha}
\left(-dw^2+d\alpha^2+\cosh^2\alpha\,dt^2\right).
\end{equation}
Now we take the world-sheet coordinates $\sigma$ and $\tau$ and the
ansatz
\begin{equation}
w=w(\sigma)\,,\qquad
\alpha=\alpha(\sigma)\,,\qquad
t=\tau\,.
\end{equation}
This ansatz is consistent and leads to the action
\begin{equation}
\cS=\frac{\sqrt\lambda}{4\pi}\int d\sigma\,d\tau\,
\frac{1}{\sinh^2\alpha}
\left(-w'^2+\alpha'^2+\cosh^2\alpha\right).
\end{equation}
Clearly $w$ is cyclic and setting it to a constant will solve the
equations of motion.

The Virasoro constraint is
\begin{equation}
\alpha'^2=\cosh^2\alpha\,,
\end{equation}
and is solved (up to a trivial shift of $\sigma$) by
\begin{equation}
\tanh\alpha=\sin\sigma\,,\qquad
0\leq\sigma<\frac{\pi}{2}\,.
\end{equation}
It is easy to check that this also solves the equation of motion
of $\alpha$, and indeed is a parametrization of the surface
$z^2=x_0^2-x_1^2-1$.

An alternative way of finding this solution (as was done for the circle in 
\cite{Beren-Corr}) is starting with
the solution for the straight line along the $x_1$ direction
at $x_0=0$, which will simply span the $z$ direction and
act on it with the $AdS_5$ isometry dual to the boost in the
$x_0$ direction
\begin{equation}
x_0\to\frac{1+x_1^2+z^2}{1-x_1^2-z^2}\,,\qquad
x_1\to\frac{2x_1}{1-x_1^2-z^2}\,,\qquad
z\to\frac{2z}{1-x_1^2-z^2}\,.
\end{equation}
This clearly is the same surface satisfying the constraint
(\ref{line-constraint}) parametrized in a different way than
above.

Yet another alternative to finding this solution are the 
techniques presented in \cite{Ishizeki:2008dw}.

The classical action for this solution diverges and requires regularization. 
The area is
\be
\cS_\text{cl.}=\frac{\sqrt\lambda}{2\pi}\int d\sigma\,d\tau\,\frac{1}{\sin^2\sigma}
=\frac{\sqrt\lambda}{2\pi}\int d\tau\left[\frac{1}{z_\text{min}}-\frac{1}{z_\text{max}}\right].
\label{zmin-max}
\ee
$z_\text{min}$ is a cutoff near the boundary of space and is usually discarded. 
$z_\text{max}$ is the maximal value of $z$ on the world-sheet and if 
we take $z_\text{max}\to\infty$, we get that the classical action vanishes.

Another possible prescription is to impose a cutoff at fixed $x_0$. Then 
ignoring the divergence from $z_\text{min}$ in \eqn{zmin-max} we have
\be
\cS_\text{cl.}=-\frac{\sqrt\lambda}{2\pi}\int d\tau
\frac{\cosh\tau}{\sqrt{x_0^2-\cosh^2\tau}}
=-\frac{\sqrt\lambda}{2}\,,
\ee
which is half of the result for the circle. Thus depending on the choice of 
cutoff we have the same result as for half the circle or the line. This is not 
surprising, since the hyperbolic line is conformal to a segment on the line or 
circle.

Since we are dealing with a supersymmetric Wilson loop it seems 
more natural, though, to follow the prescription of 
\cite{DGO}, adding a total derivative term
\be
\cS_\text{cl.}=\frac{\sqrt\lambda}{2\pi}\int d\sigma\,d\tau
\left(\frac{1}{\sin^2\sigma}+\left(\frac{z'}{z}\right)'\,\right)
=\frac{\sqrt\lambda}{2\pi}\int d\sigma\,d\tau\,\frac{1}{\cos^2\sigma}
=\frac{\sqrt\lambda}{2\pi}\int d\tau\,z_\text{max}
\ee
This expression does not diverge in the UV, near $z\sim0$, but it does 
diverge in the IR, for large $z$. Again, a possible regularization prescription 
is to integrate the world-sheet up to a cutoff $x_0=\cosh T$. This gives
\be
\frac{\sqrt\lambda}{2\pi}\int d\tau\frac{\sqrt{x_0^2-\cosh^2\tau}}{\cosh\tau}
=\frac{\sqrt\lambda}{2}(\cosh T-1)\,.
\label{line-reg1}
\ee
This is proportional to the area on $\bH_2$ of half a circle of radius $\zeta=\tanh\frac{T}{2}$, 
which also appears in the gauge theory result \eqn{bessel}. Note though, that 
the result in \eqn{bessel} corresponded to closed curves, while the hyperbolic line 
is not. Furthermore the extrapolation of \eqn{bessel} to large $\lambda$ 
gives an oscillating function, while the result we find in the $AdS$ calculation is 
real.

Another alternative is to simply keep a fixed IR cutoff $z_\text{max}$ which gives
\be
\frac{\sqrt\lambda}{2\pi}\,2Tz_\text{max}\,.
\label{line-reg2}
\ee
Note that the string world-sheet is open at future infinity (as is the dual 
Wilson loop operator), so there are different 
ways to add total derivatives which contribute differently in the infra-red. 
Furthermore, the string solution itself depends on a choice of behavior at future infinity, 
not only at the boundary at $z\to0$. The solution \eqn{line-constraint} should be 
the only one preserving $1/2$ of the supercharges, though these supercharges 
are probably broken by the cutoff $T$. It is not altogether obvious how to match 
regularization prescriptions in string theory and the gauge theory that break 
the supersymmetry in a similar way.

Some issues related to this, and to the existence of other solutions with 
different behavior at future infinity will be presented elsewhere.

Since this loop is conformal to the line, all the known descriptions of that Wilson 
loop could be adapted to this case. They include the D3-brane and D5-brane 
descriptions appropriate for loops in representations of dimension of order $N$ 
\cite{Rey-Yee, Dru-Fiol-giant, Yamaguchi:2006tq,
Gomis:2006sb, Hartnoll:2006is, Gomis:2006im} 
or the full back-reacted ``bubbling geometries'' appropriate for loops in representations 
of dimension of order $N^2$ \cite{Yamaguchi:2006te,Lunin:2006xr,
D'Hoker:2007fq, Okuda:2008px}. Likewise it is possible to calculate 
the correlation function of this Wilson loop with chiral primary operators in 
all the different pictures \cite{Beren-Corr, Giombi:2006de,Gomis:2008qa}.

\subsection{$1/4$ BPS hyperbolic cusp}
\label{subsec-cusp}

A geodesic on $\bH_2$, as mentioned before, is $1/2$-BPS and
is the analog of a great circle on $S^2$. Any two such lines will
share $1/4$ of the supercharges. Of course on $S^2$ any
two great circles will cross, while on $\bH_2$ there are
non-intersecting geodesics. Non-intersecting circles exist on $S^3$,
for example the Hopf-fibers, which indeed were an interesting example
studied in \cite{DGRT-more,DGRT-big}
.

But the intersecting circles were also interesting, since one
can then make a closed loop out of two half-circles and it is
$1/4$-BPS, which is the ``longitudes'' example in 
\cite{DGRT-more,DGRT-big} (see also
\cite{Bassetto:2008yf,Young:2008ed}). 
Now instead we can consider two hyperbolic rays meeting at a point
\begin{equation}
x^\mu=
\begin{cases}
(\cosh t,\,\sinh t,\,0,\,0)\,,&\qquad t<0\,,\cr
(\cosh t,\,-\cos\delta\sinh t,\,\sin\delta\sinh t,\,0)\,,&\qquad t>0\,.
\end{cases}
\label{cusp-loop}
\end{equation}
The scalar couplings are
\begin{equation}
\omega_i=
\begin{cases}
(1,\,0,\,0)\,dt\,,&\qquad t<0\,,\cr
(-\cos\delta,\,\sin\delta,\,0)\,dt\,,&\qquad t>0\,.
\end{cases}
\label{cusp-omega}
\end{equation}

\begin{figure}
\begin{center}
\includegraphics[width=70mm]{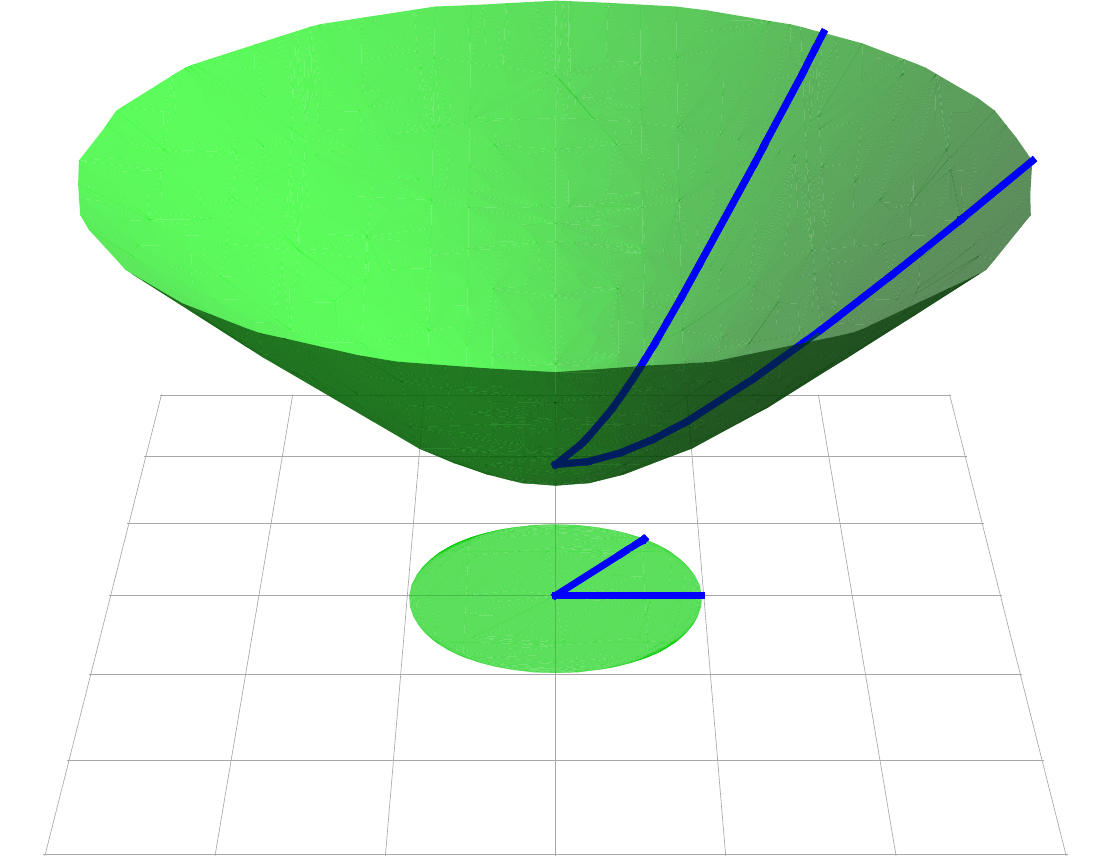}
\parbox{13cm}{
\caption{
A cusp made of two hyperbolic rays and its projection to the Poincar\'e disc.
\label{cusp-fig}}}
\end{center}
\end{figure}

As mentioned above, a hyperbolic line is conformal to a straight line,
and thus this ``hyperbolic-cusp'' is conformal to a cusp in the
$(x^1,x^2)$ plane. Explicitly
(\ref{cusp-loop}) can be written in terms of the complex
coordinates on the unit disc through (\ref{stereo}) as
\begin{equation}
\zeta=
\begin{cases}
\displaystyle \tanh\frac{t}{2}\,,&\qquad t<0\,,\cr
\vbox{\vskip8mm}
\displaystyle e^{i(\pi-\delta)}\tanh\frac{t}{2}\,,&\qquad t>0\,.
\end{cases}
\end{equation}
The Wilson loop made from this cusp in the $(x^1, x^2)$ plane and the
above scalar couplings will be also BPS, it is in the class of
operators constructed in \cite{Zarembo:2002an}. The string
solution describing this loop was written down in
\cite{DGRT-big} and it can be mapped then to the desired
configuration by the $AdS_5$ isometry which extends the
conformal transformation (\ref{stereo}) to the bulk.

We repeat here the calculation of the string surface, using the
Polyakov action and the conformal gauge rather than the
Nambu-Goto action as in \cite{DGRT-big}.

Starting with $AdS_4\times S^1$ with metric
\begin{equation}
ds^2=\frac{1}{z^2}\left(dz^2-dx_0^2+dx_1^2+dx_2^2\right)
+d\varphi^2\,,
\label{H3-metric}
\end{equation}
we change coordinates to
\begin{equation}
x_0=e^w\coth \mu\,,\qquad
x_1+ix_2 =r\,e^{i\phi}=\frac{e^{w+i\phi}\cos\nu}{\sinh \mu}\,,\qquad
z=\frac{e^w\sin\nu}{\sinh \mu}\,,
\end{equation}
so the metric becomes
\begin{equation}
ds^2=\frac{1}{\sin^2\nu}
\left(d\mu^2-\sinh^2\mu\,dw^2+d\nu^2+\cos^2\nu\,d\phi^2\right)+d\varphi^2\,.
\end{equation}
The boundary conditions for the string are at $w=0$ and it is a consistent 
ansatz to set $w=0$ along the entire world-sheet. This corresponds to the fact that 
the string is contained within an $\bH_3\times S^1$ subspace of 
$AdS_5\times S^5$ given by $x_0^2-r^2-z^2=1$.

We take now the ansatz
\begin{equation}
w=0\,,\qquad
\mu=\mu(\tau)\,,\qquad
\nu=\nu(\sigma)\,,\qquad
\phi=\phi(\sigma)\,,\qquad
\varphi=\varphi(\sigma)\,.
\end{equation}
Using dot for $\partial_\tau$ and prime for $\partial_\sigma$, the action for a 
Euclidean string world-sheet is
\begin{equation}
\cS=\frac{\sqrt\lambda}{4\pi}\int d\tau\,d\sigma
\left(\frac{1}{\sin^2\nu}
\left(\dot \mu^2+\nu'^2+\cos^2\nu\,\phi'^2\right)+\varphi'^2\right).
\end{equation}
Since the only $\tau$ dependence is in $\dot \mu$, it has to be
a constant $p$. Beyond that there are two obvious conserved
quantities
\begin{equation}
E=\cot^2\nu\,\phi'\,,\qquad
J=\varphi'\,,
\end{equation}
where we chose the names since one is related to motion on $AdS$
and the other on the sphere, but not too much should be read into
that choice of symbols.

Lastly there is the Virasoro constraint
\begin{equation}
p^2=\nu'^2+\sin^2\nu\,(E^2\tan^2\nu+J^2)\,.
\end{equation}
These sets of equations can be solved for general $E$, $J$ and
$p$, but they are particularly simple in the BPS case, when
$E^2=J^2=1-p^2$, which will turn out to be the relevant case
for us (we take them all positive). We find
\begin{equation}
\nu'^2=1-\frac{1-p^2}{\cos^2\nu}\,.
\label{nu-equation}
\end{equation}
$\nu$ varies from zero at the boundary of $AdS$ to a maximal 
value $\sin\nu=p$, at which point $\nu'=0$ and then it turns 
back towards the boundary.

The equation for $\nu$ \eqn{nu-equation} integrates to
\begin{equation}
\sin\nu=p\,\sin\sigma\,.
\end{equation}
Using the conservation equations
\begin{equation}
\phi'=\sqrt{1-p^2}\tan^2\nu
=\sqrt{1-p^2}\frac{p^2\sin^2\sigma}{1-p^2\sin^2\sigma}\,,
\qquad
\varphi'=\sqrt{1-p^2}\,,
\end{equation}
we can integrate $\phi$ and $\varphi$
\begin{align}
\tan(\phi+\sqrt{1-p^2}\,\sigma)=\sqrt{1-p^2}\tan\sigma\,,
\qquad
\varphi&=\sqrt{1-p^2}\,\sigma\,.
\end{align}

Going back to the original coordinates \eqn{H3-metric} we have
\begin{equation}
x_0=\coth p\tau\,,\qquad
x_1+ix_2=\frac{e^{-i\sqrt{1-p^2}\,\sigma}(\cos\sigma+i\sqrt{1-p^2}\sin\sigma)}
{\sinh p\tau}\,,\qquad
z=\frac{p\sin\sigma}{\sinh p\tau}\,.
\end{equation}
This solution approaches the boundary at $\sigma=0$, where $\phi=0$ and 
at $\sigma=\pi$ where $\phi=(1-\sqrt{1-p^2})\pi$. This means that $p$ is 
related to the opening angle $\delta$ in \eqn{cusp-loop} by
\be
p=\frac{1}{\pi}\sqrt{\delta(2\pi-\delta)}\,.
\ee

We wish to evaluate the action of the classical string solution. Integrating the 
area gives a divergent result, proportional to the length of the curve $2T$. 
As discussed in the previous section, there are different possible 
regularizations. Here we follow \cite{DGO} and add a total derivative term
\be
\cS_\text{cl.}=\frac{\sqrt\lambda}{2\pi}\int d\sigma\,d\tau
\left(\frac{1}{\sin^2\sigma}+\frac{z''z-z'^2+\ddot zz-\dot z^2}{z^2}\right)
=\frac{\sqrt\lambda}{2\pi}\int d\sigma\,p\,(\coth p\tau_\text{min}-1)\,.
\ee
The total derivative eliminated the divergence from the small $z$ region, but 
the integral is still divergent and it is left to specify the boundary of the 
world-sheet along which the $\sigma$ integral is to be performed.

One regularization is to consider the world-sheet up to a fixed value of 
$x_0=\cosh T=\coth p\tau_\text{min}$ giving
\be
\frac{\sqrt\lambda}{2}p\, (\cosh T-1)
=\frac{\sqrt\lambda}{2\pi}\sqrt{\delta(2\pi-\delta)}\, (\cosh T-1)\,.
\ee
This is the analog of \eqn{line-reg1}, and again is proportional to the 
area bounded by the cusp and the cutoff on $x_0$.

An alternative regularization is to restrict the world-sheet to 
$z\leq z_\text{max}$ and $x_0\leq\cosh T$. Now the range of 
integration is a bit more complicated. For 
$0\leq\sin\sigma\leq z_\text{max}/p\sinh T$ we take 
$\coth p\tau_\text{min}=\cosh T$, while in the rest of the interval we need 
to restrict to 
$\coth p\tau_\text{min}=\sqrt{z_\text{max}^2+p^2\sin^2\sigma}/p\sin\sigma$.
This gives
\begin{align}
\cS_\text{cl.}
&=\frac{\sqrt\lambda}{2\pi}\left[2p\cosh T\arcsin\frac{z_\text{max}}{p\sinh T}
-\pi p+\int d\sigma
\frac{\sqrt{z_\text{max}^2+p^2\sin^2\sigma}}{\sin\sigma}\right]
\\&\hskip-4mm
=\frac{\sqrt\lambda}{\pi}\Bigg[p\cosh T\arcsin\frac{z_\text{max}}{p\sinh T}
-p\arcsin\frac{z_\text{max}\coth T}{\sqrt{p^2+z_\text{max}^2}}
+z_\text{max}\arccosh\frac{p\cosh T}{\sqrt{p^2+z_\text{max}^2}}
\Bigg].
\nn
\end{align}

\subsection{Constant curvature curves: $1/4$ BPS circle}
\label{subsec-circle}

\begin{figure}
\begin{center}
\includegraphics[width=70mm]{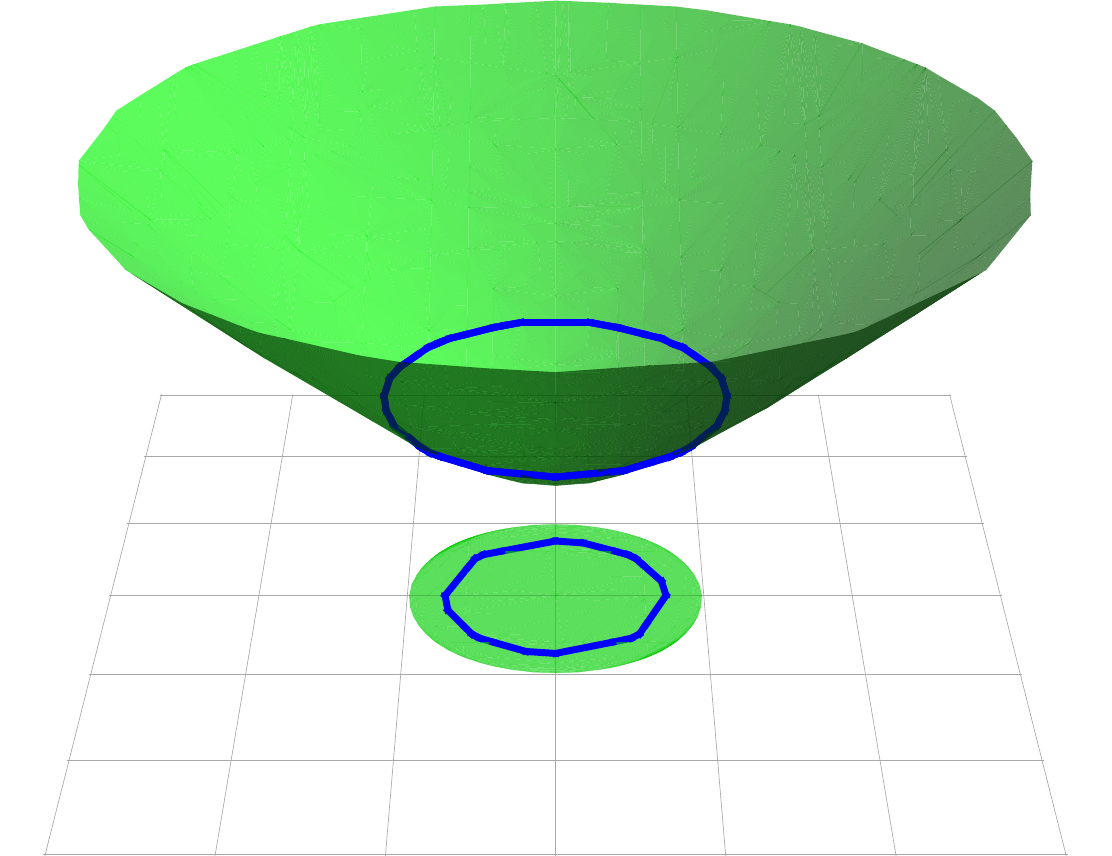}
\parbox{13cm}{
\caption{
A circle on the hyperboloid and its projection to the Poincar\'e disc.
\label{circle-fig}}}
\end{center}
\end{figure}

A special class of curves on $\bH_2$ are those with
constant curvature. The analog curves on a sphere are latitudes,
while on hyperbolic spaces there are two possibilities: Circles
and non-compact lines.

A circle is given by constant $x_0$
\begin{equation}
x=(\cosh v_0,\,\sinh v_0\cos t,\,\sinh v_0\sin t,\,0)\,,
\qquad
0\leq t\leq2\pi\,.
\label{x-circle}
\end{equation}
Our prescription \eqn{one-forms} gives periodic scalar couplings
\begin{equation}
\omega_i=\sinh v_0(-\cosh v_0\sin t,\,\cosh v_0\cos t,\,
i\sinh v_0)\,dt\,.
\label{y-circle}
\end{equation}
This loop preserves 8 supercharges, just like the latitude on
$S^2$, as can be easily verified by studying (\ref{susy}), 
(the details can be found in \cite{Volker-thesis}).

In perturbation theory this loop is a lot like the latitude on 
$S^2$ \cite{Dru-1/4}, which in turn is a lot like the usual circle 
\cite{Erickson}. The combined gauge-field 
plus scalar propagator \eqn{one-loop} is a constant
\begin{equation}
\vev{(iA_\mu^a dx^\mu+\Phi^{i\,a}\omega_i)
(iA_\mu^b dx'^\mu+\Phi^{i\,b}\omega'_i)}
=-\frac{g_{4d}^2\,\delta^{ab}}{8\pi^2}\sinh^2v_0\,dt\,dt'\,.
\end{equation}
Likewise the interaction graphs at order $g_{4d}^2$ cancel in the 
Feynman gauge. So one would expect that as in the other cases, 
the entire perturbative series will be captured by the Gaussian matrix model, 
which in the planar approximation gives \eqn{bessel}
\be
\vev{W}
\simeq\frac{2}{\sqrt{\lambda}\sinh v_0}\,
J_1\left(\sqrt{\lambda}\sinh v_0\right).
\label{circle-J1}
\ee

We turn now to finding the string dual of this Wilson loop.
Unlike the previous two examples, the line and cusp, in this case the 
scalar couplings \eqn{y-circle} are complex. Therefore we will not 
be able to embed the string dual in $AdS_5\times S^5$, rather we 
need to consider a complexification of this space.

\begin{figure}
\begin{center}
\includegraphics[width=150mm]{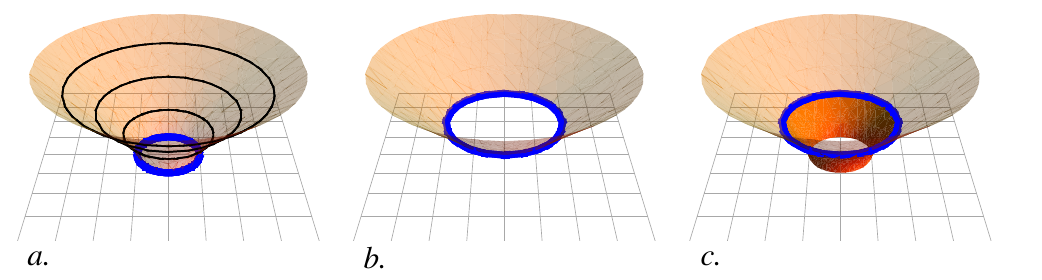}
\parbox{13cm}{
\caption{
A representation of the minimal surface solutions for the circle on the 
hyperboloid. ($a.$) The $AdS$ metric is Wick-rotated to $dS_3$ where the string 
fills half of a $dS_2$ subspace bounded by the blue circle at $z=0$. The 
$S^2\subset S^5$ is Wick-rotated to $dS_2$ where the boundary values 
are again represented by the blue circle. The surface either covers the 
part of space above the circle ($b.$) or below the circle ($c.$) where the string is 
represented on the orbifold $dS_2/\bZ_2$, so the surface is reflected back 
up from the $y_0=0$ plane.
\label{circlesurf-fig}}}
\end{center}
\end{figure}

In the case of the latitude the string solution is given in the metric \eqn{S2-metric} 
by \cite{Dru-Fiol-int,Dru-1/4}
\be
\begin{gathered}
x_1+ix_2=\frac{e^{i\tau}\tanh\sigma_0}{\cosh\sigma}\,,\qquad
x_3=\frac{1}{\cosh\sigma_0}\,,\qquad
z=\tanh\sigma_0\tanh\sigma\,,\\
y_1+iy_2
=\frac{e^{i\tau}}{\cosh(\sigma_0\pm\sigma)}\,,\qquad
y_3
=\tanh(\sigma_0\pm\sigma)\,.
\end{gathered}
\label{latitude}
\ee
This loop ends at the boundary at $\sigma=0$ along the latitude 
at $|x_1+ix_2|=\tanh\sigma_0$ and $x_3=1/\cosh\sigma_0$.

Now we want to consider instead a loop ending along the circle 
with $|x_1+ix_2|>1$, which can be represented by an imaginary 
$\sigma_0$. We therefore Wick-rotate both $\sigma$ and $\sigma_0$ in 
the latitude solution into
\be
\begin{gathered}
x_1+ix_2=\frac{e^{i\tau}\tan\sigma_0}{\cos\sigma}\,,\qquad
x_0=\frac{1}{\cos\sigma_0}\,,\qquad
z=\tan\sigma_0\tan\sigma\,,\\
y_1+iy_2=\frac{e^{i\tau}}{\cos(\sigma_0\pm\sigma)}\,,\qquad
y_0=\tan(\sigma_0\pm\sigma)\,.
\end{gathered}
\label{circle-solution}
\ee
While the original solution satisfied
\be
z^2+x_1^2+x_2^2+x_3^2=1\,,\qquad
y_1^2+y_2^2+y_3^2=1\,,
\ee
this new configuration satisfies
\be
z^2+x_0^2-x_1^2-x_2^2=1\,,\qquad
y_1^2+y_2^2-y_0^2=1\,.
\ee
This indeed fits a target space Wick-rotated to the metric \eqn{dS-metric}.

One can easily check that \eqn{circle-solution} is a solution of the equations of 
motion for a string with Lorentzian world-sheet in the metric \eqn{dS-metric}
\be
\begin{aligned}
\cS=\frac{\sqrt\lambda}{4\pi}\int d\tau\,d\sigma\bigg(
\frac{z'^2+x_0'^2-x_1'^2-x_2'^2+x_1^2+x_2^2}{z^2}
&-y_0'^2+y_1'^2+y_2'^2-y_1^2-y_2^2
\\&+\Lambda(y_1^2+y_2^2-y_0^2-1)\bigg).
\end{aligned}
\ee
On the classical solution this is equal to
\begin{equation}
\cS_\text{cl.}=\sqrt{\lambda}\int d\sigma\left(
\frac{1}{\sin^2\sigma}-\frac{1}{\cos^2(\sigma_0\pm\sigma)}\right).
\end{equation}
We have not specified the range of $\sigma$ integration. $\sigma=0$ is the 
boundary of space, where we should impose $y_0=\tan\sigma_0=\sinh v_0$. 
There is the usual UV divergence, which can be removed by adding a total 
derivative $(z''z-z'^2)/z^2$. At $\sigma\to\pi/2$ the solution reaches 
$|x_1+ix_2|\to\infty$, which 
is a reasonable end for that part of the solution. For the $y_0$ coordinate 
it seems like we should also allow it to go to infinity, corresponding to 
$\sigma=\pi/2\mp\sigma_0$. Then we have
\begin{equation}
\cS_\text{cl.}=\sqrt\lambda\int d\sigma\left(
\frac{1}{\cos^2\sigma}-\frac{1}{\cos^2(\sigma_0\pm\sigma)}\right)
=\sqrt\lambda\big(\tan\sigma\mp\tan(\sigma_0\pm\sigma)\big).
\end{equation}
With this peculiar choice of boundary conditions the divergences at 
$\sigma=\pi/2$ for the first term and from $\sigma=\pi/2\mp\sigma_0$ for the 
second term cancel and we end up with the contribution from $\sigma=0$
\begin{equation}
\cS_\text{cl.}=\pm\sqrt{\lambda}\tan\sigma_0
=\pm\sqrt{\lambda}\sinh v_0\,.
\end{equation}

The choice of sign corresponds to two solutions one where $y_0\to\infty$ and 
the other where $y_0\to-\infty$. Together we get that the expectation value of this 
Wilson loop behaves like
\be
\vev{W_\text{circle}}\sim\sum_\pm e^{i\cS_\text{cl.}}
\sim\cos\left(\sqrt{\lambda}\sinh v_0\right).
\ee
This agrees with the matrix model result \eqn{circle-J1}.

It should be possible to find the D3-brane dual to this Wilson loop in analogy to 
the calculation in \cite{DGRT-1/4-giant} as well as its coupling to chiral 
primary operators \cite{Semenoff:2006am}.

\subsection{Constant curvature curves: $1/4$ BPS hyperbolic line}
\label{subsec-1/4-line}

\begin{figure}
\begin{center}
\includegraphics[width=70mm]{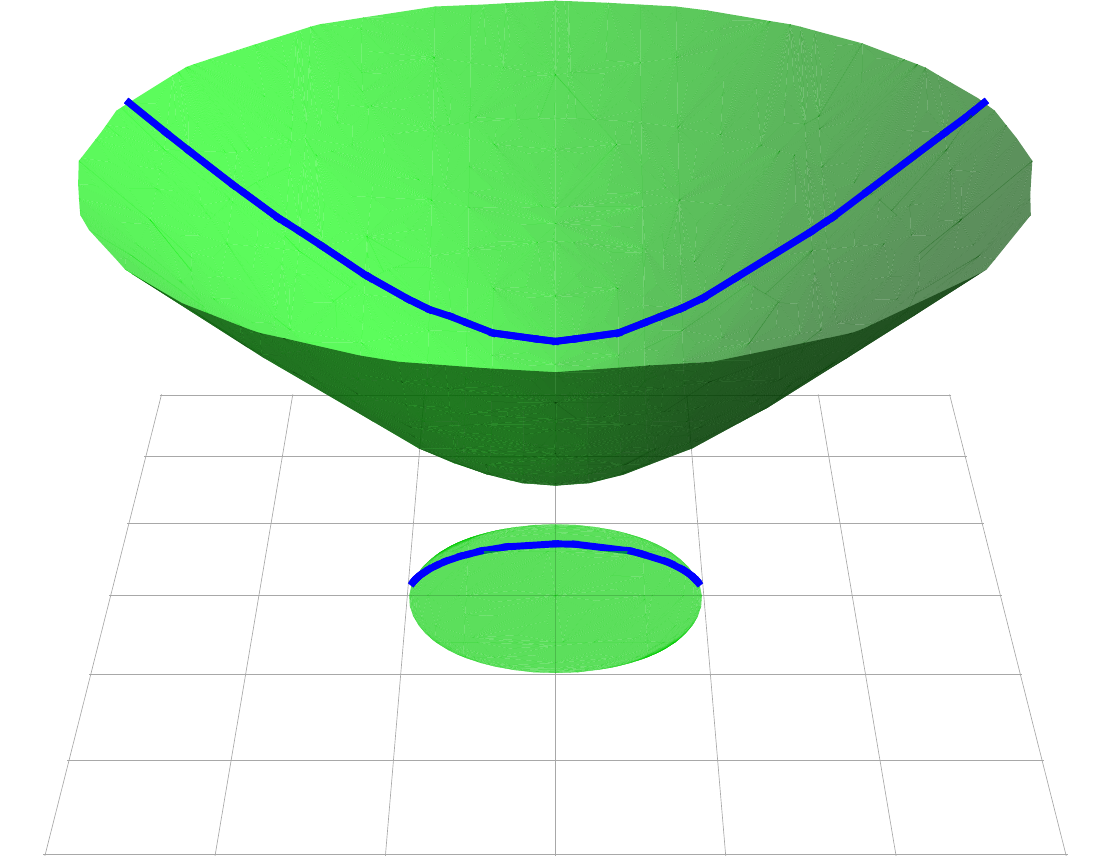}
\parbox{13cm}{
\caption{
A constant curvature hyperbolic line at $x_2=1$ on the hyperboloid 
and its stereographic projection onto the Poincar\'e disc.
\label{quarter-fig}}}
\end{center}
\end{figure}

The other class of constant curvature curves on $\bH_2$ is
that of hyperbolic lines whose curvature does not match that of
the underlying space. Such a line is given by
\begin{equation}
x=(\cosh v_0\cosh t,\,\cosh v_0\sinh t,\,\sinh v_0,\,0)\,,
\qquad
-\infty<t<\infty\,,
\end{equation}
and the scalar couplings are now hyperbolic
\begin{equation}
\omega_i=\cosh v_0(\cosh v_0,\,-\sinh v_0\sinh t,\,
-i\sinh v_0\cosh t)\,dt\,.
\end{equation}
Like the previous example, this loop also preserves 8 supercharges. 
And yet again, the combined propagator is a constant
\begin{equation}
\vev{(iA_\mu^a dx^\mu+\Phi^{i\,a}\omega_i)
(iA_\mu^b dx'^\mu+\Phi^{i\,b}\omega'_i)}
=-\frac{g_{4d}^2\,\delta^{ab}}{8\pi^2}\cosh^2v_0\,dt\,dt'\,.
\end{equation}
Unlike the previous example, though, it is not easy to sum up ladder diagrams. 
This Wilson loop is non-compact and the calculation diverges at both 
ends of the line. As the situation in the simpler case of the 1/2 BPS 
hyperbolic line is complicated enough, we do not try to resolve the 
issues associated with these divergences here.

\begin{figure}
\begin{center}
\includegraphics[width=150mm]{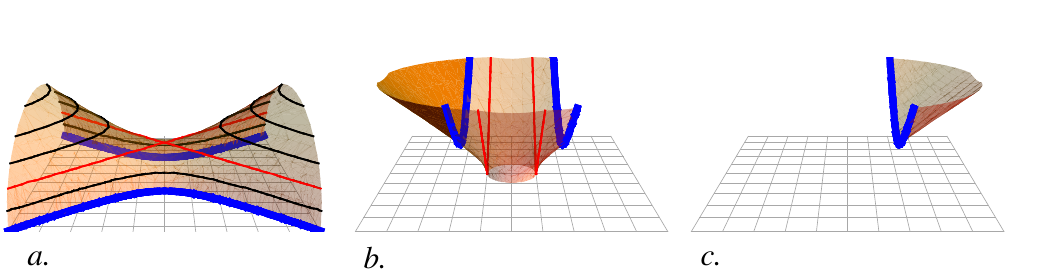}
\parbox{14cm}{
\caption{
A representation of the minimal surface solutions for the $1/4$ BPS hyperbolic 
line. ($a.$) The string ends on two lines 
on the hyperboloids with positive and negative $x_0$, but one can consider 
only half of the world-sheet, assuming the string dual lives on $dS_3/\bZ_2$. 
The red dotted lines are the borders between 
the coordinate patches described in the text. 
For the correlator of two lines, the string should end also along two hyperbolic 
lines on $dS_2$ ($b.$) and the string goes to one side and is reflected 
from infinity back. If considering just a single line, it should be enough to take 
only the piece of the string going to the right as in ($c.$), or to the left. 
\label{quartersurf-fig}}}
\end{center}
\end{figure}

In considering the string dual we take the solution of the latitude on $S^2$ 
\eqn{latitude} and analytically continue $\tau\to i\tau$ and $\sigma_0\to\sigma_0-i\pi/2$. 
This gives
\be
\begin{gathered}
x_0=\frac{\cosh\tau\coth\sigma_0}{\cosh\sigma}\,,\quad\ 
x_1=\frac{\sinh\tau\coth\sigma_0}{\cosh\sigma}\,,\quad\ 
x_2=\frac{1}{\sinh\sigma_0}\,,\quad\ 
z=\coth\sigma_0\tanh\sigma\,,\\
y_1=\coth(\sigma_0\pm\sigma)\,,\qquad
y_2=-\frac{\sinh\tau}{\sinh(\sigma_0\pm\sigma)}\,,\qquad
y_0=\frac{\cosh\tau}{\sinh(\sigma_0\pm\sigma)}\,.
\end{gathered}
\label{1/4-line-1}
\ee
where we have relabeled the coordinates and Wick rotated three of them. This 
solution is now embedded in a target-space with metric \eqn{dS-metric}.

The boundary conditions are satisfied for $\sinh\sigma_0=1/\sinh v_0$ and 
it is easy to check that this is a solution of the equations of motion for a 
Lorentzian world-sheet.
As $\sigma\to\infty$ the world-sheet approaches the curve given by
$z=\cosh v_0$, $x_2=\sinh v_0$ and $x_0=x_1$. The world-sheet has to be 
analytically continued and we find another patch of the solution with
\be
\begin{gathered}
x_0=\frac{\sinh\tau\coth\sigma_0}{\sinh\sigma}\,,\quad\ 
x_1=\frac{\cosh\tau\coth\sigma_0}{\sinh\sigma}\,,\quad\ 
x_2=\frac{1}{\sinh\sigma_0}\,,\quad\ 
z=\coth\sigma_0\coth\sigma\,,\\
y_1=\tanh(\sigma_0\pm\sigma)\,,\qquad
y_2=\frac{\cosh\tau}{\cosh(\sigma_0\pm\sigma)}\,,\qquad
y_0=\frac{\sinh\tau}{\cosh(\sigma_0\pm\sigma)}\,.
\end{gathered}
\label{1/4-line-2}
\ee
Finally we need to add a third patch to the world-sheet which is identical to 
\eqn{1/4-line-1} only with negative $x_0$ and $y_1$.

The resulting string solution ends along two curves on the boundary. One on the 
original hyperboloid with $x_0>0$ and the other on the second hyperboloid with 
$x_0<0$. This has been forced on us through the analytical continuation of the 
solution. We note though that perhaps it is legitimate to consider only half 
of this solution. With the Wick-rotation of the coordinate $z$ the $AdS$ space has 
turned into $dS$ and away from $x_0=z=0$ we can do the same $\bZ_2$ identification 
on this space as we did on the de-Sitter space that replaced $S^2$. Then the 
solution will be given by one patch like \eqn{1/4-line-1} and half of \eqn{1/4-line-2}, 
with $x_0\geq0$.

\section*{Acknowledgments}
We would like to thank Bartomeu Fiol, Juan Maldacena and Jan Plefka 
for stimulating discussions. 
N.D would like to thank The Galileo Galilei Institute in Florence, 
the Tata Institute for Fundamental Research, The Niels Bohr Institute and the 
Technion for their welcoming hospitality 
in the course of this work and the INFN and ICTS for partial financial support.


\begin{thebibliography}{10}
\addtolength{\parskip}{-1ex}

\bibitem{D'Hoker:2008ix}
E.~D'Hoker, J.~Estes, M.~Gutperle, D.~Krym, and P.~Sorba, ``{Half-BPS
  supergravity solutions and superalgebras},''
  \href{http://dx.doi.org/10.1088/1126-6708/2008/12/047}{{\em JHEP} {\bf 12}
  (2008)  047},
\href{http://arxiv.org/abs/0810.1484}{{\tt arXiv:0810.1484}}.

\bibitem{DGRT-more}
N.~Drukker, S.~Giombi, R.~Ricci, and D.~Trancanelli, ``{More supersymmetric
  Wilson loops},'' \href{http://dx.doi.org/10.1103/PhysRevD.76.107703}{{\em
  Phys. Rev.} {\bf D76} (2007)  107703},
\href{http://arxiv.org/abs/0704.2237}{{\tt arXiv:0704.2237}}.

\bibitem{DGRT-YM2}
N.~Drukker, S.~Giombi, R.~Ricci, and D.~Trancanelli, ``{Wilson loops: From
  four-dimensional SYM to two-dimensional YM},''
  \href{http://dx.doi.org/10.1103/PhysRevD.77.047901}{{\em Phys. Rev.} {\bf
  D77} (2008)  047901},
\href{http://arxiv.org/abs/0707.2699}{{\tt arXiv:0707.2699}}.

\bibitem{DGRT-big}
N.~Drukker, S.~Giombi, R.~Ricci, and D.~Trancanelli, ``{Supersymmetric Wilson
  loops on $S^3$},''
  \href{http://dx.doi.org/10.1088/1126-6708/2008/05/017}{{\em JHEP} {\bf 05}
  (2008)  017},
\href{http://arxiv.org/abs/0711.3226}{{\tt arXiv:0711.3226}}.

\bibitem{Zarembo:2002an}
K.~Zarembo, ``{Supersymmetric Wilson loops},''
  \href{http://dx.doi.org/10.1016/S0550-3213(02)00693-4}{{\em Nucl. Phys.} {\bf
  B643} (2002)  157--171},
\href{http://arxiv.org/abs/hep-th/0205160}{{\tt hep-th/0205160}}.

\bibitem{Guralnik:2003di}
Z.~Guralnik and B.~Kulik, ``{Properties of chiral Wilson loops},'' {\em JHEP}
  {\bf 01} (2004)  065,
\href{http://arxiv.org/abs/hep-th/0309118}{{\tt hep-th/0309118}}.

\bibitem{Guralnik:2004yc}
Z.~Guralnik, S.~Kovacs, and B.~Kulik, ``{Less is more: Non-renormalization
  theorems from lower dimensional superspace},'' {\em Int. J. Mod. Phys.} {\bf
  A20} (2005)  4546--4553,
\href{http://arxiv.org/abs/hep-th/0409091}{{\tt hep-th/0409091}}.

\bibitem{Dymarsky:2006ve}
A.~Dymarsky, S.~S. Gubser, Z.~Guralnik, and J.~M. Maldacena, ``{Calibrated
  surfaces and supersymmetric Wilson loops},'' {\em JHEP} {\bf 09} (2006)  057,
\href{http://arxiv.org/abs/hep-th/0604058}{{\tt hep-th/0604058}}.

\bibitem{Erickson}
J.~K. Erickson, G.~W. Semenoff, and K.~Zarembo, ``{Wilson loops in ${\cal N} =
  4$ supersymmetric Yang-Mills theory},''
  \href{http://dx.doi.org/10.1016/S0550-3213(00)00300-X}{{\em Nucl. Phys.} {\bf
  B582} (2000)  155--175},
\href{http://arxiv.org/abs/hep-th/0003055}{{\tt hep-th/0003055}}.

\bibitem{Dru-Gross}
N.~Drukker and D.~J. Gross, ``{An exact prediction of ${\cal N} = 4$ SUSYM
  theory for string theory},'' \href{http://dx.doi.org/10.1063/1.1372177}{{\em
  J. Math. Phys.} {\bf 42} (2001)  2896--2914},
\href{http://arxiv.org/abs/hep-th/0010274}{{\tt hep-th/0010274}}.

\bibitem{Pestun:2007rz}
V.~Pestun, ``{Localization of gauge theory on a four-sphere and supersymmetric
  Wilson loops},''
\href{http://arxiv.org/abs/0712.2824}{{\tt arXiv:0712.2824}}.

\bibitem{Dru-1/4}
N.~Drukker, ``{1/4 BPS circular loops, unstable world-sheet instantons and the
  matrix model},'' \href{http://dx.doi.org/10.1088/1126-6708/2006/09/004}{{\em
  JHEP} {\bf 09} (2006)  004},
\href{http://arxiv.org/abs/hep-th/0605151}{{\tt hep-th/0605151}}.

\bibitem{Bassetto:2008yf}
A.~Bassetto, L.~Griguolo, F.~Pucci, and D.~Seminara, ``{Supersymmetric Wilson
  loops at two loops},''
  \href{http://dx.doi.org/10.1088/1126-6708/2008/06/083}{{\em JHEP} {\bf 06}
  (2008)  083},
\href{http://arxiv.org/abs/0804.3973}{{\tt arXiv:0804.3973}}.

\bibitem{Young:2008ed}
D.~Young, ``{BPS Wilson loops on $S^2$ at higher loops},''
  \href{http://dx.doi.org/10.1088/1126-6708/2008/05/077}{{\em JHEP} {\bf 05}
  (2008)  077},
\href{http://arxiv.org/abs/0804.4098}{{\tt arXiv:0804.4098}}.

\bibitem{Ishizeki:2008dw}
R.~Ishizeki, M.~Kruczenski, and A.~Tirziu, ``{New open string solutions in
  $AdS_5$},'' \href{http://dx.doi.org/10.1103/PhysRevD.77.126018}{{\em Phys.
  Rev.} {\bf D77} (2008)  126018},
\href{http://arxiv.org/abs/0804.3438}{{\tt arXiv:0804.3438}}.

\bibitem{Alday:2007hr}
L.~F. Alday and J.~M. Maldacena, ``{Gluon scattering amplitudes at strong
  coupling},'' {\em JHEP} {\bf 06} (2007)  064,
\href{http://arxiv.org/abs/0705.0303}{{\tt arXiv:0705.0303}}.

\bibitem{Drummond:2007aua}
J.~M. Drummond, G.~P. Korchemsky, and E.~Sokatchev, ``{Conformal properties of
  four-gluon planar amplitudes and Wilson loops},''
  \href{http://dx.doi.org/10.1016/j.nuclphysb.2007.11.041}{{\em Nucl. Phys.}
  {\bf B795} (2008)  385--408},
\href{http://arxiv.org/abs/0707.0243}{{\tt arXiv:0707.0243}}.

\bibitem{Brandhuber:2007yx}
A.~Brandhuber, P.~Heslop, and G.~Travaglini, ``{MHV amplitudes in ${\cal N}=4$
  super Yang-Mills and Wilson loops},''
  \href{http://dx.doi.org/10.1016/j.nuclphysb.2007.11.002}{{\em Nucl. Phys.}
  {\bf B794} (2008)  231--243},
\href{http://arxiv.org/abs/0707.1153}{{\tt arXiv:0707.1153}}.

\bibitem{Bern:2008ap}
Z.~Bern {\em et al.}, ``{The two-loop six-gluon MHV amplitude in maximally
  supersymmetric Yang-Mills theory},''
  \href{http://dx.doi.org/10.1103/PhysRevD.78.045007}{{\em Phys. Rev.} {\bf
  D78} (2008)  045007},
\href{http://arxiv.org/abs/0803.1465}{{\tt arXiv:0803.1465}}.

\bibitem{Drummond:2008aq}
J.~M. Drummond, J.~Henn, G.~P. Korchemsky, and E.~Sokatchev, ``{Hexagon Wilson
  loop = six-gluon MHV amplitude},''
\href{http://arxiv.org/abs/0803.1466}{{\tt arXiv:0803.1466}}.

\bibitem{Alday:2008yw}
L.~F. Alday and R.~Roiban, ``{Scattering Amplitudes, Wilson Loops and the
  String/Gauge Theory Correspondence},''
  \href{http://dx.doi.org/10.1016/j.physrep.2008.08.002}{{\em Phys. Rept.} {\bf
  468} (2008)  153--211},
\href{http://arxiv.org/abs/0807.1889}{{\tt arXiv:0807.1889}}.

\bibitem{Volker-thesis}
V.~Branding, ``{Supersymmetric Wilson loops in the $AdS$/CFT
  correspondence},''. Diploma thesis, Humboldt University Berlin, 2008;
  available at http://qft.physik.hu-berlin.de.

\bibitem{DGRT-1/4-giant}
N.~Drukker, S.~Giombi, R.~Ricci, and D.~Trancanelli, ``{On the D3-brane
  description of some 1/4 BPS Wilson loops},''
  \href{http://dx.doi.org/10.1088/1126-6708/2007/04/008}{{\em JHEP} {\bf 04}
  (2007)  008},
\href{http://arxiv.org/abs/hep-th/0612168}{{\tt hep-th/0612168}}.

\bibitem{Wu:1977hi}
T.~T. Wu, ``{Two-dimensional Yang-Mills theory in the leading $1/N$
  expansion},''
\href{http://dx.doi.org/10.1016/0370-2693(77)90762-6}{{\em Phys. Lett.} {\bf
  B71} (1977)  142}.

\bibitem{Mandelstam:1982cb}
S.~Mandelstam, ``{Light cone superspace and the ultraviolet finiteness of the
  ${\cal N}=4$ model},''
\href{http://dx.doi.org/10.1016/0550-3213(83)90179-7}{{\em Nucl. Phys.} {\bf
  B213} (1983)  149--168}.

\bibitem{Leibbrandt:1983pj}
G.~Leibbrandt, ``{The light cone gauge in Yang-Mills theory},''
\href{http://dx.doi.org/10.1103/PhysRevD.29.1699}{{\em Phys. Rev.} {\bf D29}
  (1984)  1699}.

\bibitem{Staudacher:1997kn}
M.~Staudacher and W.~Krauth, ``{Two-dimensional {QCD} in the
  Wu-Mandelstam-Leibbrandt prescription},''
  \href{http://dx.doi.org/10.1103/PhysRevD.57.2456}{{\em Phys. Rev.} {\bf D57}
  (1998)  2456--2459},
\href{http://arxiv.org/abs/hep-th/9709101}{{\tt hep-th/9709101}}.

\bibitem{Bassetto:1998sr}
A.~Bassetto and L.~Griguolo, ``{Two-dimensional {QCD}, instanton contributions
  and the perturbative Wu-Mandelstam-Leibbrandt prescription},''
  \href{http://dx.doi.org/10.1016/S0370-2693(98)01319-7}{{\em Phys. Lett.} {\bf
  B443} (1998)  325--330},
\href{http://arxiv.org/abs/hep-th/9806037}{{\tt hep-th/9806037}}.

\bibitem{Witten:1992xu}
E.~Witten, ``{Two-dimensional gauge theories revisited},''
  \href{http://dx.doi.org/10.1016/0393-0440(92)90034-X}{{\em J. Geom. Phys.}
  {\bf 9} (1992)  303--368},
\href{http://arxiv.org/abs/hep-th/9204083}{{\tt hep-th/9204083}}.

\bibitem{Rey-Yee}
S.-J. Rey and J.-T. Yee, ``{Macroscopic strings as heavy quarks in large $N$
  gauge theory and anti-de Sitter supergravity},''
  \href{http://dx.doi.org/10.1007/s100520100799}{{\em Eur. Phys. J.} {\bf C22}
  (2001)  379--394},
\href{http://arxiv.org/abs/hep-th/9803001}{{\tt hep-th/9803001}}.

\bibitem{Maldacena-wl}
J.~M. Maldacena, ``{Wilson loops in large $N$ field theories},''
  \href{http://dx.doi.org/10.1103/PhysRevLett.80.4859}{{\em Phys. Rev. Lett.}
  {\bf 80} (1998)  4859--4862},
\href{http://arxiv.org/abs/hep-th/9803002}{{\tt hep-th/9803002}}.

\bibitem{BMN}
D.~E. Berenstein, J.~M. Maldacena, and H.~S. Nastase, ``{Strings in flat space
  and pp waves from ${\cal N}=4$ Super Yang Mills},''
\href{http://dx.doi.org/10.1063/1.1524550}{{\em AIP Conf. Proc.} {\bf 646}
  (2003)  3--14}.

\bibitem{Yoneya:2003mu}
T.~Yoneya, ``{What is holography in the plane-wave limit of $AdS_5$/SYM$_4$
  correspondence?},'' \href{http://dx.doi.org/10.1143/PTPS.152.108}{{\em Prog.
  Theor. Phys. Suppl.} {\bf 152} (2004)  108--120},
\href{http://arxiv.org/abs/hep-th/0304183}{{\tt hep-th/0304183}}.

\bibitem{Strominger:2001pn}
A.~Strominger, ``{The $dS$/CFT correspondence},'' {\em JHEP} {\bf 10} (2001)
  034,
\href{http://arxiv.org/abs/hep-th/0106113}{{\tt hep-th/0106113}}.

\bibitem{Balasubramanian:2001nb}
V.~Balasubramanian, J.~de~Boer, and D.~Minic, ``{Mass, entropy and holography
  in asymptotically de Sitter spaces},''
  \href{http://dx.doi.org/10.1103/PhysRevD.65.123508}{{\em Phys. Rev.} {\bf
  D65} (2002)  123508},
\href{http://arxiv.org/abs/hep-th/0110108}{{\tt hep-th/0110108}}.

\bibitem{Maldacena:2002vr}
J.~M. Maldacena, ``{Non-Gaussian features of primordial fluctuations in single
  field inflationary models},'' {\em JHEP} {\bf 05} (2003)  013,
\href{http://arxiv.org/abs/astro-ph/0210603}{{\tt astro-ph/0210603}}.

\bibitem{Schroedinger}
E.~Schr\"odinger, {\em Expanding universes}.
\newblock University Press, Cambridge [Eng.], 1956.

\bibitem{Parikh:2002py}
M.~K. Parikh, I.~Savonije, and E.~P. Verlinde, ``{Elliptic de Sitter space:
  $dS/Z_2$},'' \href{http://dx.doi.org/10.1103/PhysRevD.67.064005}{{\em Phys.
  Rev.} {\bf D67} (2003)  064005},
\href{http://arxiv.org/abs/hep-th/0209120}{{\tt arXiv:hep-th/0209120}}.

\bibitem{Beren-Corr}
D.~E. Berenstein, R.~Corrado, W.~Fischler, and J.~M. Maldacena, ``{The operator
  product expansion for Wilson loops and surfaces in the large $N$ limit},''
  \href{http://dx.doi.org/10.1103/PhysRevD.59.105023}{{\em Phys. Rev.} {\bf
  D59} (1999)  105023},
\href{http://arxiv.org/abs/hep-th/9809188}{{\tt hep-th/9809188}}.

\bibitem{DGO}
N.~Drukker, D.~J. Gross, and H.~Ooguri, ``{Wilson loops and minimal
  surfaces},'' \href{http://dx.doi.org/10.1103/PhysRevD.60.125006}{{\em Phys.
  Rev.} {\bf D60} (1999)  125006},
\href{http://arxiv.org/abs/hep-th/9904191}{{\tt hep-th/9904191}}.

\bibitem{Dru-Fiol-giant}
N.~Drukker and B.~Fiol, ``{All-genus calculation of Wilson loops using
  D-branes},'' \href{http://dx.doi.org/10.1088/1126-6708/2005/02/010}{{\em
  JHEP} {\bf 02} (2005)  010},
\href{http://arxiv.org/abs/hep-th/0501109}{{\tt hep-th/0501109}}.

\bibitem{Yamaguchi:2006tq}
S.~Yamaguchi, ``{Wilson loops of anti-symmetric representation and
  D5-branes},'' \href{http://dx.doi.org/10.1088/1126-6708/2006/05/037}{{\em
  JHEP} {\bf 05} (2006)  037},
\href{http://arxiv.org/abs/hep-th/0603208}{{\tt hep-th/0603208}}.

\bibitem{Gomis:2006sb}
J.~Gomis and F.~Passerini, ``{Holographic Wilson loops},'' {\em JHEP} {\bf 08}
  (2006)  074,
\href{http://arxiv.org/abs/hep-th/0604007}{{\tt hep-th/0604007}}.

\bibitem{Hartnoll:2006is}
S.~A. Hartnoll and S.~P. Kumar, ``{Higher rank Wilson loops from a matrix
  model},'' {\em JHEP} {\bf 08} (2006)  026,
\href{http://arxiv.org/abs/hep-th/0605027}{{\tt hep-th/0605027}}.

\bibitem{Gomis:2006im}
J.~Gomis and F.~Passerini, ``{Wilson loops as D3-branes},'' {\em JHEP} {\bf 01}
  (2007)  097,
\href{http://arxiv.org/abs/hep-th/0612022}{{\tt hep-th/0612022}}.

\bibitem{Yamaguchi:2006te}
S.~Yamaguchi, ``{Bubbling geometries for half BPS Wilson lines},''
  \href{http://dx.doi.org/10.1142/S0217751X07035070}{{\em Int. J. Mod. Phys.}
  {\bf A22} (2007)  1353--1374},
\href{http://arxiv.org/abs/hep-th/0601089}{{\tt hep-th/0601089}}.

\bibitem{Lunin:2006xr}
O.~Lunin, ``{On gravitational description of Wilson lines},''
  \href{http://dx.doi.org/10.1088/1126-6708/2006/06/026}{{\em JHEP} {\bf 06}
  (2006)  026},
\href{http://arxiv.org/abs/hep-th/0604133}{{\tt hep-th/0604133}}.

\bibitem{D'Hoker:2007fq}
E.~D'Hoker, J.~Estes, and M.~Gutperle, ``{Gravity duals of half-BPS Wilson
  loops},'' \href{http://dx.doi.org/10.1088/1126-6708/2007/06/063}{{\em JHEP}
  {\bf 06} (2007)  063},
\href{http://arxiv.org/abs/0705.1004}{{\tt arXiv:0705.1004}}.

\bibitem{Okuda:2008px}
T.~Okuda and D.~Trancanelli, ``{Spectral curves, emergent geometry, and
  bubbling solutions for Wilson loops},''
  \href{http://dx.doi.org/10.1088/1126-6708/2008/09/050}{{\em JHEP} {\bf 09}
  (2008)  050},
\href{http://arxiv.org/abs/0806.4191}{{\tt arXiv:0806.4191}}.

\bibitem{Giombi:2006de}
S.~Giombi, R.~Ricci, and D.~Trancanelli, ``{Operator product expansion of
  higher rank Wilson loops from D-branes and matrix models},''
  \href{http://dx.doi.org/10.1088/1126-6708/2006/10/045}{{\em JHEP} {\bf 10}
  (2006)  045},
\href{http://arxiv.org/abs/hep-th/0608077}{{\tt hep-th/0608077}}.

\bibitem{Gomis:2008qa}
J.~Gomis, S.~Matsuura, T.~Okuda, and D.~Trancanelli, ``{Wilson loop correlators
  at strong coupling: from matrices to bubbling geometries},''
  \href{http://dx.doi.org/10.1088/1126-6708/2008/08/068}{{\em JHEP} {\bf 08}
  (2008)  068},
\href{http://arxiv.org/abs/0807.3330}{{\tt arXiv:0807.3330}}.

\bibitem{Dru-Fiol-int}
N.~Drukker and B.~Fiol, ``{On the integrability of Wilson loops in $AdS_5\times
  S^5$: Some periodic ans\"atze},''
  \href{http://dx.doi.org/10.1088/1126-6708/2006/01/056}{{\em JHEP} {\bf 01}
  (2006)  056},
\href{http://arxiv.org/abs/hep-th/0506058}{{\tt hep-th/0506058}}.

\bibitem{Semenoff:2006am}
G.~W. Semenoff and D.~Young, ``{Exact 1/4 BPS loop: Chiral primary
  correlator},'' \href{http://dx.doi.org/10.1016/j.physletb.2006.10.047}{{\em
  Phys. Lett.} {\bf B643} (2006)  195--204},
\href{http://arxiv.org/abs/hep-th/0609158}{{\tt hep-th/0609158}}.

\end{thebibliography}

\providecommand{\href}[2]{#2}\begingroup\raggedright\endgroup

\end{document}